# Superconductivity enhancement in polar metal regions of $Sr_{0.95}Ba_{0.05}TiO_3$ and $Sr_{0.985}Ca_{0.015}TiO_3$ revealed by systematic Nb doping

Yasuhide Tomioka, Naoki Shirakawa and Isao H. Inoue[*]

National Institute of Advanced Industrial Science and Technology (AIST),

Tsukuba 305-8565, Japan

---

[*]email: isaocaius@gmail.com

Two different ferroelectric materials, $Sr_{0.95}Ba_{0.05}TiO_3$ and $Sr_{0.985}Ca_{0.015}TiO_3$, can be turned into polar metals with broken centrosymmetry via electron doping. Systematic substitution of $Nb^{5+}$ for $Ti^{4+}$ has revealed that these polar metals both commonly show a simple superconducting dome with a single convex shape. Interestingly, the superconducting transition temperature $T_c$ is enhanced more strongly in these polar metals when compared with the nonpolar matrix $Sr(Ti,Nb)O_3$. The maximum $T_c$ reaches 0.75 K, which is the highest reported value among the $SrTiO_3$-based families to date. However, the $T_c$ enhancement is unexpectedly lower within the vicinity of the putative ferroelectric quantum critical point. The enhancement then becomes much more prominent at locations further inside the dilute carrier-density region, where the screening is less effective. These results suggest that centrosymmetry breaking, i.e., the ferroelectric nature, does not kill the superconductivity. Instead, it enhances the superconductivity directly, despite the absence of strong quantum fluctuations.

**Keywords:** $SrTiO_3$, Ba susbstitution, ferroelectricity, Nb doping, polar metal, superconducting dome, superconductivity enhancement

Are superconductivity and ferroelectricity reconcilable? Bernd T. Matthias, a pioneer in the search for superconducting materials, stated more than 50 years ago[1] that "superconductivity and ferroelectricity will exclude one another." This is known as the Matthias conjecture, which is an old guideline used in the hunt for new superconducting materials. The conflict between superconductivity and ferroelectricity motivated the study of Bednorz and Müller in which they discovered high-temperature cuprate superconductors[2]. Anderson and Blount proposed that some metals can be ferroelectric in the sense that they have broken inversion symmetry. However, almost all ferroelectric materials are highly insulating, and only a few of these materials have been experimentally shown to be Anderson-Blount-type ferroelectric (polar) metals[3]. These so-called polar metals have recently raised interest from the research community[4–21]. In particular, low-carrier-density polar metals have drawn attention[22–25] because they show superconductivity that contradicts the Matthias conjecture. Our study focuses on one of these metals, the perovskite-type titanate $SrTiO_3$, to reveal a missing aspect of superconductivity with ferroelectricity.

Single crystals of $SrTiO_3$ show huge relative permittivity of approximately 42000 at 4.2 K (ref.[26]). However, the ferroelectric phase transition does not occur at temperatures above 35 mK (ref.[27]). Because the ground state is predicted to be ferroelectric[28,29], it is believed that strong quantum fluctuations suppress the material's ferroelectricity[27]. However, this does not explain the unusual minimum observed in the inverse relative permittivity of $SrTiO_3$ (refs.[30,31]), and the origin of the paraelectric state is still a subject of intensive debate. Regardless of its origin, the paraelectric state is quite fragile; a small amount of homovalent substitution of $Ca^{2+}$ (refs.[32,33]) or $Ba^{2+}$ (refs.[34,35]) for $Sr^{2+}$ turns $SrTiO_3$ into a ferroelectric material. Exchange of the $^{18}O$ isotope for $^{16}O$ also results in a change into the ferroelectric phase (refs.[10,36]). The domain walls of $SrTiO_3$ in the tetragonal phase carry polar properties related to ferroelectricity[37,38]. Furthermore, both epitaxial strain[18,19,39,40] and Sr defects[41] in thin $SrTiO_3$ films also generate ferroelectricity.

In contrast, electron doping of SrTiO$_3$, e.g., substitution of La$^{3+}$ or Sm$^{3+}$ for Sr$^{2+}$ (refs.[4–7,42,43]) or Nb$^{5+}$ for Ti$^{4+}$ (refs.[21,44–48]), or removal of O$^{2-}$ (refs.[9–12,49–54]), causes SrTiO$_3$ to become metallic. In all these cases, the superconductivity appears at low temperatures. (Note that carrier doping using Nb or La is denoted by Nb:SrTiO$_3$ or La:SrTiO$_3$ in this paper in cases where it is not necessary to express the compositions explicitly as SrTi$_{1-x}$Nb$_x$O$_3$ or Sr$_{1-x}$La$_x$TiO$_3$.) A recent tunnelling study revealed that the Ti 3$d$ $t_{2g}$ orbitals generate two light bands and one heavy band[48]. The electrons in the heavy band make the largest contribution to the superconductivity[48]. The most interesting property of doped SrTiO$_3$ systems is that they are extremely low carrier density superconductors with Fermi energies that are a few orders of magnitude lower than the phonon energy. Because the Migdal-Eliashberg criterion (adiabatic condition) is violated[23,24], the pairing interaction should show a significant frequency dependence that is caused by poor screening of the Coulomb repulsion[24]. Further doping causes the Fermi energy to increase, but the superconductivity disappears above a carrier density of 10$^{21}$ cm$^{-3}$, and this results in a dome-like dependence of the superconducting transition temperature on the carrier density[4,49–51].

The physical mechanism that lies behind these phenomena has challenged the related microscopic theory for more than half a century. Although no compelling theories have been proposed that encompass the disparate existing ideas, some unconventional mechanisms have recently sparked new research interest in this field. Typical examples include the Cooper pairings driven by plasmon and plasmon-polariton coupling with longitudinal optical (LO) phonons[55–57] and by ferroelectric fluctuations[15–17,22,24,29], particularly those driven via the two phonons[58–61] of a soft transverse optical (TO) mode, which explain the $T^2$-dependence of the resistivity very well[62,63]. There were also some remarkable experimental discoveries: carrier doping of two ferroelectric matrices, (Sr,Ca)TiO$_3$ (refs.[10–12]) and SrTi($^{16}$O,$^{18}$O)$_3$ (refs.[4,9,10]), was found to raise the transition temperature $T_c$ to be higher than that of the SrTiO$_3$ matrix, indicating that the superconductivity appears to be enhanced in the ferroelectric SrTiO$_3$. Several experimental investigations of SrTiO$_3$ with epitaxial strain were also noteworthy[5–7,43]; the strain breaks the material's spatial inversion symmetry and enhances the spin-orbit interactions[5,6], and $T_c$ then increased to 0.6 K (ref.[5]). Surprisingly, although $T_c$ is very low in these SrTiO$_3$ systems, magnetic impurities, e.g. Sm, Eu, demonstrated no effect on $T_c$ (ref.[43]).

We have studied two different polar metals, comprising Sr$_{0.95}$Ba$_{0.05}$TiO$_3$ and Sr$_{0.985}$Ca$_{0.015}$TiO$_3$, with substitution of Nb$^{5+}$ for Ti$^{4+}$ for carrier densities ranging from ~10$^{18}$ to ~10$^{21}$ cm$^{-3}$. Higher levels of Ca or Ba substitution may extend the ferroelectric-metal region of the carrier density to be much greater. However, Ba substitution at more than 20% is accompanied by a complex sequential structural change, and Ca substitution at levels above 0.9% has not been studied in depth. Therefore, 5% for Ba and 1.5% for Ca are acceptably good settings. Sr$_{0.95}$Ba$_{0.05}$TiO$_3$ exhibits Ti-site-dominated ferroelectricity, in which Ti-O hybridisation causes a strong pseudo- or second-order Jahn-Teller distortion[13]. The Curie temperature is ~50 K when the polarisation direction is along the [111] direction of a pseudo-cubic lattice. In contrast, Sr$_{0.985}$Ca$_{0.015}$TiO$_3$ shows Sr-site-dominated ferroelectricity that originates from the off-centre position of Ca$^{2+}$, which has a smaller ionic radius than Sr$^{2+}$ (refs.[11–14]). The Curie temperature is ~30 K when the polarisation direction is [110] (ref.[32]). Despite their differences in terms of ferroelectricity, we demonstrate that these two polar metals show common and simple superconducting domes with a higher $T_c$ than that of the nonpolar matrix Nb:SrTiO$_3$. We also show that the $T_c$ enhancement becomes much greater if we go deeper into the polar region.

**Results**

**Nonpolar matrix Nb:SrTiO$_3$ compared to SrTiO$_{3-\delta}$ and La:SrTiO$_3$.** The evolution of the metallic state of our Nb:SrTiO$_3$ single crystals (see the Methods for details of our sample preparation processes) is illustrated in Fig. 1a. The results are similar to those reported by Tufte et al.[64] Hall effect measurements (see Supplementary Note 3) were performed to deduce the carrier density $n$ and the Hall mobility $\mu$ at 5 K (Fig. 1b). The values of $\mu$ for SrTiO$_{3-\delta}$ and Nb:SrTiO$_3$ from the literature[44,64–66] are also plotted for comparison. In the normal metal, $\mu$ is proportional to $n^{-1}$, but Behnia proposed a model[67] that assumed that the mean-free path is proportional to the average distance between the dopants and the Thomas-Fermi screening length, thus giving $\mu \propto n^{-\frac{5}{6}}$. Therefore, we tried to fit the formula $\mu \propto n^{-b}$ to our experimental data (see Supplementary Figure 2). The result that $b = 0.85 \pm 0.02$ indicates that $b = 5/6$ from the literature[67] falls within the error bar, whereas $b = 1$ does not. In fact, our data in Fig. 1b fitted the model reasonably well. In the model, the proportionality factor is dependent on the reciprocal of the effective mass and the dopant potential. Because the $\mu$ value of our Nb:SrTiO$_3$ is greater than that of SrTiO$_{3-\delta}$, the dopant potential is shallower for Nb:SrTiO$_3$. Creation of oxygen defects causes one or two electrons to be trapped at each oxygen-vacancy site and localised without contributing to the itinerant carriers[67,68]. However, in the case of Ti/Nb substitution, the replacement of the Ti 3$d$ orbitals with the Nb 4$d$ orbitals does not change the orbital characteristics, which means that the disorder is smaller in scale than that caused by the formation of oxygen defects. This is expected to increase the superconducting critical temperature because the spatial disorder generally destroys the superconducting state and suppresses $T_c$ (ref.[69]).

The resistive transitions of the superconductivity of our Nb:SrTiO$_3$ single crystals (i.e. the same samples studied in Fig. 1a) are shown in Fig. 2a. We defined $T_c$ as the mid-point of the resistive transition, as described in ref.[4] and as shown in Fig. 2**a**. The upper and lower error bars were determined from the onset and end temperatures of the superconductivity. This definition of $T_c$ is used throughout this work. The $T_c$ values are plotted as a function of the carrier density $n$ at 5 K in Fig. 2**b**. The carrier density in the nonpolar metallic state shows little dependence on temperature, as described in the Supplementary Note 2. The $T_c$ values of SrTiO$_{3-\delta}$ (refs.[12,49]), Nb:SrTiO$_3$ (ref.[23]) and La:SrTi($^{16}$O$_{1-z}$$^{18}$O$_z$)$_3$ (refs.[4,42] and the three unpublished data points depicted in Supplementary Figure 1) are also plotted for comparison. Our data clearly demonstrate that the superconducting dome of Nb:SrTiO$_3$ is shifted toward a higher $T_c$ and a larger $n$ region than the superconducting dome of La:SrTiO$_3$. The optimal $T_c$ value for our Nb:SrTiO$_3$ is ~0.5 K at $n \simeq 1\times 10^{20}$ cm$^{-3}$.

There are two noticeable differences between the superconducting domes of some of the SrTiO$_{3-\delta}$ samples in the literature and the domes of our La:SrTiO$_3$ and Nb:SrTiO$_3$ samples. The first prominent difference is a shoulder peak that occurs at $n \simeq 1.2\times 10^{18}$ cm$^{-3}$ in SrTiO$_{3-\delta}$ (ref.[12]). If the three-fold degeneracy of the $t_{2g}$ band is lifted[51] and each band has a different filling, it is then likely that a superconducting dome will be observed for each band. However, neither our La:SrTiO$_3$ nor our Nb:SrTiO$_3$ samples showed this shoulder peak, seemingly indicating that a single band of Ti 3$d$ makes the contribution to the superconductivity[48]. The second difference is observed in the high $n$ region ($n \simeq 1\times 10^{21}$ cm$^{-3}$), where superconductivity with $T_c \simeq 0.25$ K was reported for SrTiO$_{3-\delta}$ (ref.[49]). Similar to the previous case, neither our La:SrTiO$_3$ sample nor our Nb:SrTiO$_3$ sample showed superconductivity. One possible explanation that can account for both differences simultaneously is based on consideration of the inhomogeneity of the oxygen vacancies[52]. The thermal reduction procedure used to create the oxygen vacancies for the SrTiO$_{3-\delta}$ single crystal is restricted to dislocation[70] because the formation enthalpy of the oxygen vacancies near the dislocations is significantly lower than that in the stoichiometric SrTiO$_3$ matrix[71]. Therefore, the inhomogeneous carrier density distribution occurs in the SrTiO$_{3-\delta}$ sample. Hall measurements give the averaged carrier density of the bulk, whereas the observed



value of $T_c$ is the highest $T_c$ among all the percolation paths in the doped regions where the carrier density differs from the average value.

When compared with La:SrTiO$_3$ and SrTiO$_{3-\delta}$, Nb:SrTiO$_3$ shows either comparable or higher mobility (Fig. 1**b**; see also Supplementary Fig. 3 of ref.[4]) and a higher $T_c$ with a simple single superconducting dome (Fig. 2**b**). These results mean that Nb doping of the ferroelectric derivatives of SrTiO$_3$ represents a solid strategy for investigation of the relationship between ferroelectricity and superconductivity. Therefore, we focused on the two ferroelectric matrices: Sr$_{0.985}$Ca$_{0.015}$TiO$_3$ and Sr$_{0.95}$Ba$_{0.05}$TiO$_3$. Interestingly, the types of ferroelectricity exhibited by these two matrices are different.

**Polar metals Nb:(Sr,Ca)TiO$_3$ and Nb:(Sr,Ba)TiO$_3$.** In (Sr,Ca)TiO$_3$, the ferroelectricity is driven by dipole-dipole interactions between the off-centre Ca sites, which have smaller ionic radii than the Sr$^{2+}$ ion[13]. Nb doping makes this material highly conductive, but its resistivity increases slowly as $T$ decreases at low temperatures (Fig. 3**a**). In contrast, our nonpolar Nb:SrTiO$_3$ single crystal does not show this resistance anomaly at all. We therefore defined $T_K$ as the temperature at which the resistivity reaches a minimum. The values of $T_K$ were plotted versus $n$ at $T_K$ ($n$ at 5 K for $T_K$ = 0) and fitted the solid line in Fig. 3**b** fairly well. To derive $n$ at $T_K$, we performed a smoothing spline interpolation[72] (smoothing factor = 1) for the raw $n$ vs. $T$ data (see Supplementary Note 10) using Igor Pro v8.04 software (WaveMetrics, Inc., USA). Intriguingly, the ferroelectric Curie temperature $T_C^{\mathrm{FE}}$ ~25 K, corresponding to the sharp peak in the relative permittivity $\varepsilon$ (see the inset of Fig. 3**b**), is located almost on the same line. It was proposed that the dipole moment remains at the off-centre Ca site in the lightly carrier-doped (Sr,Ca)TiO$_3$ because of poor screening, and the glassy dipole-dipole interaction between the Ca sites then causes the resistance anomaly[12,14]. The static interaction is fully screened ($T_K$ = 0) at a value of $n^*$ of 3.1×10$^{19}$ cm$^{-3}$. This experimental value of $n^*$ is almost equivalent to the $n^*$ = 3.3×10$^{19}$ cm$^{-3}$ value calculated by assuming a Thomas-Fermi screening length for the itinerant carriers that is comparable with the averaged dipole-dipole distance[14].



In contrast, in (Sr,Ba)TiO$_3$, the Ba$^{2+}$ ion hardly moves because of its larger ionic radius and heavier mass when compared with the Sr$^{2+}$ ion. Therefore, phonon softening through the pseudo- or second-order Jahn-Teller distortion that occurs because of the sizeable Ti-O hybridisation becomes the central mechanism of the ferroelectricity[13]. This is different in principle from the ferroelectricity mechanism in (Sr,Ca)TiO$_3$, but is basically similar to that in SrTi($^{18}$O,$^{16}$O)$_3$, in which the Ti-O hybridisation is also crucial. Our Sr$_{0.95}$Ba$_{0.05}$TiO$_3$ single crystal shows the ferroelectric transition at $T_\text{C}^\text{FE}$ ~50 K (inset of Fig. 3**d**), which is accompanied by a structural change from cubic *P*m3m to rhombohedral *R*3m (ref.[34]). We performed powder X-ray diffraction measurements on our metallic Sr$_{0.95}$Ba$_{0.05}$Ti$_{0.998}$Nb$_{0.002}$O$_3$ at 10 K, and the results were examined via a Rietveld analysis (see Supplementary Note 11). The pattern was, in fact, consistent with the rhombohedral *R*3m symmetry that occurs with displacement of Ti along the [111] axis of a pseudo-cubic lattice. The spatial inversion symmetry is broken and we thus consider the metallic state to be polar. To our surprise, although the origin of its ferroelectricity differs from that of Nb:Sr$_{0.985}$Ca$_{0.015}$TiO$_3$, a similar resistance anomaly is observed in Nb:Sr$_{0.95}$Ba$_{0.05}$TiO$_3$ (Fig. 3**c**), and the value of $T_\text{K}$ decreases with increasing *n* (Fig. 3**d**). For samples with two local minima in their resistivity characteristics, the mid-point between the two temperatures that give these minima is defined as $T_\text{K}$, where the lower and upper ends of the error bar correspond to each of these local minima. The carrier density dependence does not seem to be as linear as that observed in Nb:Sr$_{0.985}$Ca$_{0.015}$TiO$_3$, but if a linear relationship between $T_\text{K}$ and *n* is assumed simply to separate the polar and nonpolar regions, the line almost reaches $T_\text{C}^\text{FE}$ at *n* = 0. We have estimated a critical carrier density of $n^*$ ~ 2.5×10$^{20}$ cm$^{-3}$ at $T_\text{K}$ = 0. This value of $n^*$ is one order of magnitude higher than that of Nb:Sr$_{0.985}$Ca$_{0.015}$TiO$_3$, although the difference in $T_\text{C}^\text{FE}$ is only a factor of two. (Note that even if we assume that the polar/nonpolar boundary has a different shape, this does not affect the order of $n^*$.)

Russel et al. reported that the resistance anomaly temperature $T_\text{K}$ of their strained (Sr,Sm)TiO$_3$ film appeared at precisely the same temperature at which the second harmonic



generation (SHG) signal showed a sharp increase[18]. Because SHG indicates breaking of the inversion symmetry of the system, it is reasonable to define the polar metal region based on $T_K$, i.e. both Nb:Sr$_{0.985}$Ca$_{0.015}$TiO$_3$ and Nb:Sr$_{0.95}$Ba$_{0.05}$TiO$_3$ are polar metals in the shaded areas shown in Fig. 3**b** and 3**d**, respectively. Note that, at temperatures below $T_K$, the carrier density $n$ for Nb:Sr$_{0.95}$Ba$_{0.05}$TiO$_3$ decreases with decreasing temperature (see Supplementary Note 7) in a similar manner to that reported in ref.[5]. This means that some carriers are bound to the local dipole moment for screening. However, the values of $n$ above $T_K$ are not dependent on the temperature and are similar to the value of $n$ at room temperature. The critical carrier density $n^*$, at which the resistance anomaly disappears, is discussed because it may represent the putative ferroelectric quantum critical point (ref.[4,15,18]) because breaking of the centrosymmetry occurs when $n < n^*$ (ref.[18]), possibly as a result of poor screening. If so, it is quite intriguing that our two systems, i.e. Nb:Sr$_{0.985}$Ca$_{0.015}$TiO$_3$ and Nb:Sr$_{0.95}$Ba$_{0.05}$TiO$_3$, have different ferroelectric mechanisms and show one order of magnitude of difference in their $n^*$ values ($3.1 \times 10^{19}$ cm$^{-3}$ for Nb:Sr$_{0.985}$Ca$_{0.015}$TiO$_3$ and $2.5 \times 10^{20}$ cm$^{-3}$ for Nb:Sr$_{0.95}$Ba$_{0.05}$TiO$_3$). It is thus essential to consider whether these significant differences will affect the appearance of the superconductivity. From this point, we investigate the superconductivity of these two systems systematically.

**Superconductivity of Nb:(Sr,Ca)TiO$_3$ and Nb:(Sr,Ba)TiO$_3$.** Both polar metals, i.e. Nb:Sr$_{0.985}$Ca$_{0.015}$TiO$_3$ and Nb:Sr$_{0.95}$Ba$_{0.05}$TiO$_3$, show superconductivity at low temperatures. The resistive transition of the former material is shown in Fig. 4**a**, and that of the latter material is shown in Fig. 4**b**. The $T_c$ values plotted as a function of $n$ at 5 K produce a single superconducting dome, as in the case of Nb:SrTiO$_3$ (Fig. 4**c**, bottom). Moreover, the optimal $T_c$ value of Nb:Sr$_{0.95}$Ba$_{0.05}$TiO$_3$ reaches 0.75 K at $n \simeq 1 \times 10^{20}$ cm$^{-3}$. This is the highest $T_c$ value among the SrTiO$_3$ families reported to date. Additionally, another interesting point can be observed in this viewgraph. Although the types of ferroelectricity in Sr$_{0.985}$Ca$_{0.015}$TiO$_3$ and Sr$_{0.95}$Ba$_{0.05}$TiO$_3$ are different and their $n^*$ values differ by one order of magnitude, it is common that the value of $n$ for the maximum of the superconducting dome in both cases is almost identical to that for the nonpolar Nb:SrTiO$_3$, whereas their superconductivity persists even in the



extremely low carrier density region. For clarity, we performed the same smoothing spline interpolation procedure that was used to derive $n$ at $T_K$ for Fig. 3**b** and 3**d**. The dashed lines in the bottom panel of Fig. 4**c** show that the spline interpolations fitted the raw data very well. Using these interpolations, we deduced the $T_c$ difference between the polar and nonpolar systems for each value of $n$, as indicated by the solid lines in the top panel of Fig. 4**c**. We expected initially that the largest difference would be seen at around $n^*$ because we believed that the quantum fluctuation could be the main driving force of the superconductivity and that it may have become strongest at $n^*$. However, the increase in $T_c$ around $n^*$ was not significant (less than 0.1 K) for both Nb:Sr$_{0.985}$Ca$_{0.015}$TiO$_3$ and Nb:Sr$_{0.95}$Ba$_{0.05}$TiO$_3$. Surprisingly, as the system moves deeper into the polar region, in which the ferroelectric fluctuations must be suppressed, the superconductivity still persists in the two polar metals.

It is thus necessary to consider the mechanism that contributes to the considerable enhancement of the superconductivity that occurs in the dilute carrier density region away from $n^*$. One fascinating idea[7,21,54,73] is to consider the intrinsic inhomogeneity in this region. Indeed, formation of polar nanodomains is essential in most polar metals; small domains grow as the carrier density is reduced, and macroscopic ferroelectricity finally appears at the zero carrier density limit[19]. It has recently been revealed that application of uniaxial strain to the Nb:SrTiO$_3$ single crystal induces both dislocations and local ferroelectricity with inhomogeneity that result in enhancement of the material's superconductivity[21]. Electroforming also creates metallic filaments, and the resulting inhomogeneous structure increases the superconducting transition temperature[54].

**Meissner effect of Nb:(Sr,Ba)TiO$_3$.** However, we should note that our polar Nb:Sr$_{0.95}$Ba$_{0.05}$TiO$_3$ single crystal ($n \sim 8.1 \times 10^{19}$ cm$^{-3}$), which gives the highest resistive $T_c$ (0.75 K), is not in such a greatly inhomogeneous state that filaments would be formed. (For the nonpolar Nb:SrTiO$_3$, filamentary superconductivity is at least ruled out at optimum doping levels[46]). In fact, the $T_c$ value that is defined as the onset of the AC mutual inductance drop (0.70 K) is almost equal to



the corresponding resistive value (0.75 K), as illustrated in the top panel of Fig. 5, and this suggests that the $T_c$ distribution is homogeneous. In the bottom panel of Fig. 5, we plotted the DC volume susceptibility versus temperature. The onset of diamagnetism occurs at around 0.70 K, which nearly coincides with the zero resistivity temperature; this indicates that the superconductivity of our Nb:Sr$_{0.95}$Ba$_{0.05}$TiO$_3$ cannot be regarded as being at least filamentary on the nearly static timescale. Although the volume fraction of the Meissner effect (the reversible expulsion of the magnetic flux during the field cooling measurement) appears to be small, it is almost comparable to that of La:SrTiO$_3$ ($n \sim 1.6 \times 10^{20}$ cm$^{-3}$) (ref.[42]), where localised moments that were six orders of magnitude smaller than that of the La ions caused such a small volume fraction. Because these extremely small amounts of impurities never affect the $T_c$ value at all, the apparently small volume fraction is not an important issue here. (Collignon et al. measured the superfluid density of nonpolar Nb:SrTiO$_3$, which is equal to the normal state density, thus indicating that the superconductivity is a bulk phenomenon[47].) Therefore, the maximum $T_c$ is seen to be found in higher $n$ regions, where the inhomogeneity is suppressed; in contrast, in this study, we focus on $T_c$ enhancement in the lower $n$ region located far away from $n^*$, where the inhomogeneity is important.

**Discussion**

Bretz-Sullivan et al. investigated[53] single-crystalline SrTiO$_{3-\delta}$ within the dilute single band limit for $3.9 \times 10^{16}$ cm$^{-3}$ < $n$ < $1.4 \times 10^{18}$ cm$^{-3}$. Although the system held a three-dimensional homogeneous electron gas in the normal state, the superconducting state was inhomogeneous. Interestingly, the $T_c$ values for all the nonpolar SrTiO$_{3-\delta}$ samples in the inhomogeneous superconducting region are not enhanced and remain almost constant at around 65 mK. This result means that the inhomogeneous superconductivity in the dilute carrier-density region may not be the dominant cause of the $T_c$ enhancement; i.e., the ferroelectric nature of the domains would be essential to the observed enhancement. For the two-dimensional superconductivities at the LaAlO$_3$/SrTiO$_3$ interface, Rashba spin-orbit coupling is key to the Cooper pairing



phenomenon, i.e., the superconductivity is sympathetic to inversion symmetry breaking[74]. In bulk SrTiO$_3$, the spin-orbit interaction is also essential[6,17,24,48,75] and the ferroelectricity arises from breaking of the centrosymmetry[17]. Therefore, the bulk superconductivity may involve more space symmetry breaking, which goes against the Matthias conjecture. In this sense, it is intriguing that the enigmatic 2 K superconductivity observed in the (111) surface of KTaO$_3$ is discussed as being related to a large spin-orbit interaction[76]. Furthermore, based on this scenario, we may realize higher $T_c$ values by increasing the Ba content to raise $T_C^{FE}$. Indeed, recent theoretical work on carrier-doped BaTiO$_3$ has indicated that significant modulation of the electron-phonon coupling occurs across the polar-to-centrosymmetric phase transition, and superconductivity is then predicted to occur at 2 K (ref.[20]). Another theoretical work on Dirac semimetals predicted that the superconductivity will only appear in the ferroelectric region[25]. The two-phonon exchange superconductivity scenario[58–61] requires neither quantum criticality nor quantum fluctuations, but can predict a reasonable $T_c$ value in the dilute carrier-density region[60]. Therefore, we considered whether the ferroelectric fluctuations that occur around $n^*$ may not be the main driving force for enhancement of the superconductivity. However, the dynamic movement of the polar-domain boundaries can be regarded as a type of spatiotemporal fluctuation of $n^*$. In other words, the intrinsic inhomogeneity of the materials may still play some role in the $T_c$ enhancement. The lattice-polarity-superconductivity dynamics must be investigated further to provide a comprehensive overview of the unique superconductivity of SrTiO$_3$.

We have confirmed that Nb doping is an excellent carrier doping method to produce SrTiO$_3$ with reduced disorder, superior mobility, and a higher $T_c$. Sr$_{0.985}$Ca$_{0.015}$TiO$_3$ and Sr$_{0.95}$Ba$_{0.05}$TiO$_3$ are different types of ferroelectrics, and we confirmed via X-ray structural analyses that Nb doping turns these materials into polar metals. When the carrier density $n < n^*$, the resistance shows an anomaly at $T_K$, which increases monotonically as the carrier density decreases and coincides with the ferroelectric Curie temperature at the zero carrier density limit. Although the values of $n^*$ for Nb:Sr$_{0.985}$Ca$_{0.015}$TiO$_3$ and Nb:Sr$_{0.95}$Ba$_{0.05}$TiO$_3$ are different by one order of



magnitude, both materials show typical single superconducting domes with peaks commonly located around $10^{20}$ cm$^{-3}$. When compared with that of nonpolar Nb:SrTiO$_3$, the values of $T_c$ are enhanced up to 0.75 K. However, the increase in $T_c$ was not significant (less than 0.1 K) at around $n^*$. This enhancement becomes much more prominent as we go deeper into the polar metal region. Space-symmetry breaking not only coexists with superconductivity, but also enhances the superconductivity. Our results call for a reconsideration of the existing microscopic models of superconductivity in SrTiO$_3$.



**Methods**

**Single crystal growth.** For Nb:SrTiO$_3$, we mixed powders of SrCO$_3$, TiO$_2$, and Nb$_2$O$_5$ in a ratio of 1 : 1−x : x/2. For Nb:Sr$_{0.985}$Ca$_{0.015}$TiO$_3$, we mixed powders of SrCO$_3$, CaCO$_3$, TiO$_2$. and Nb$_2$O$_5$ in a ratio of 0.985 : 0.015 : 1−x : x/2. For Nb:Sr$_{0.95}$Ba$_{0.05}$TiO$_3$, we mixed powders of SrCO$_3$, BaCO$_3$, TiO$_2$, and Nb$_2$O$_5$ in a ratio of 0.95 : 0.05 : 1−x : x/2. In all these cases, the values of x multiplied by 100 correspond to the atomic % of the nominal Nb content in the sample. The powders were calcined at 700 °C in the air for a few hours. The calcined powders were sintered at 1000 °C in the air for five hours. Then, the powders were pulverised and formed into a rod, about 4 mm in diameter and about 50 mm in length. Each rod was fired at 1250 °C – 1350 °C for five hours in an argon gas flow. The crystal growth was conducted in a floating zone furnace with double hemi-ellipsoidal mirrors coated with gold. Two halogen lamps were used as the heat source. The crystals were grown in a stream of argon gas, and the growth rate was settled at 10 – 15 mm/h.

**Structural analyses.** Powder X-ray diffraction (XRD) patterns were collected on the Cu *Kα* radiation diffractometer (SmartLab, Rigaku Co., Ltd) using the *θ-2θ* step scanning method in the range of 15° ≤ 2*θ* ≤ 110°. The pattern indicated that the samples were of a single phase. For some crystals, Rietveld refinements for the lattice parameters and crystal symmetry were performed at various temperatures from 300 K to 10 K. Crystal alignments were done using back-reflection Laue diffraction. Results are summarised in Supplementary Note 11.

**Transport properties.** We cut the samples into rectangular shapes of 0.5 × 0.3 × 7 mm$^3$ with the longest edge parallel to the [100] direction of the cubic indices. Electrodes for the measurements were made by ultrasonic indium soldering, and the current was injected parallel to the [100] direction. The resistivity and the Hall voltage for 5 K ≤ *T* ≤ 300 K were measured in the Physical Property Measurement System (PPMS, Quantum Design Inc.). The resistivity



below 1 K was measured using an AC resistance bridge (LR700, Linear Research Inc.) in a cryostat using a $^3$He/$^4$He dilution refrigerator ($\mu$ dilution, Taiyo-Toyo Sanso Inc.).

**Relative permittivity.** Same as the transport measurements, the samples were cut along the [100] direction with the typical dimensions of 1.5 × 0.4 × 3 mm$^3$. The (110) plane is the widest surface on which the electrodes were formed by painting and drying the silver paste. The relative permittivity for 5 K ≤ $T$ ≤ 300 K were measured in PPMS.

**Mutual inductance.** The measurements were performed using the LR700 resistance bridge in a cryostat using a $^3$He/$^4$He dilution refrigerator ($\mu$ dilution, Taiyo-Toyo Sanso Inc.). Induction (detection) coils with the diameter/length of 3 mm / 11 mm (2 mm / 6 mm) were set directly on the single crystals. We used a superconducting NbTi wire to avoid a possible temperature rise due to the Joule heating of the coils. The excitation current was 2 mA. The amplitude of the AC magnetic field $\mu_0 H$ ($\mu_0$ is the permeability of the vacuum) was estimated to be approximately 0.02 mT. The frequency was set at 15.9 Hz. Mutual inductance is proportional to the voltage of the detection coil, and the transition temperature is defined as the onset of the mutual inductance drop.

**Diamagnetism.** The DC magnetisation measurement for the Nb:Sr$_{0.95}$Ba$_{0.05}$TiO$_3$ single crystal ($n \sim 8.1 \times 10^{19}$ cm$^{-3}$) of 1.1 × 1.1 × 7.0 mm$^3$ was performed using a superconducting quantum interference device (SQUID) magnetometer (MPMS Quantum Design Inc.) equipped with a $^3$He refrigerator (*i*Helium 3, IQUANTUM Inc.). The DC magnetic field was applied along the longest edge of the crystal parallel to the [100] direction of the pseudo-cubic lattice. The demagnetising factor along this direction is estimated to be less than 0.046 [Osborn, J. A. Demagnetizing Factors of the General Ellipsoid. *Phys. Rev.* **67**, 351–357 (1945).]; thus, we can ignore the demagnetising effect. In the zero-field cooling (ZFC) protocol, the sample temperature was lowered to 0.4 K in zero field. The DC magnetisation was measured in the presence of a static DC magnetic field $\mu_0 H$ of 0.02, 0.05, and 0.1 mT while warming the sample up to above $T_c$. In



the field cooling (FC) protocol, the sample temperature was lowered to 0.4 K in the presence of $\mu_0 H$, and the DC magnetisation was measured in the static $\mu_0 H$ while warming. There is a tiny field-independent background in magnetisation. Since the contributions from the sample to the SQUID signals are so small, the raw data are far from the ideal dipole response because of the background. We therefore subtracted a SQUID curve above $T_c$ from the curves below $T_c$ before performing the fitting.

**Data availability**

The data supporting this paper's plots and other findings of this study are available in figshare with the digital object identifier <https://doi.org/10.6084/m9.figshare.21524499>. Further data and resources are available from the corresponding authors upon reasonable request.

**Acknowledgements**

This study was supported by the Japan Society for the Promotion of Science (JSPS) KAKENHI Grant Nos. 19H01844 (Category B) and 18H03686 (Category A) and was partially supported by the Japan Science and Technology Agency (JST) CREST Grant No. JPMJCR19K2. The authors would like to thank Hisashi Inoue, Keisuke Shibuya and Reiji Kumai for helping us with the experiments, and Yaron Kedem, Jonathan Ruhman, Susanne Stemmer, Gian G. Guzmán-Verri, Siddharth (Montu) Saxena and Alessio Zaccone for stimulating discussions.

**Figure Legends**

**Fig. 1 | Resistivity and Hall mobility of Nb:SrTiO$_3$ single crystals. a**, Temperature dependence of resistivity with nominal Nb content $x$ values of 0.0002, 0.0005, 0.001, 0.002, 0.0035, 0.005, 0.007, 0.01, 0.015, 0.02 and 0.05 over the range from room temperature down to 3 K. The numbers of Ti 3$d$ electrons deduced from Hall effect measurements are noted in parentheses. **b**, Hall mobilities of our samples at 5 K (blue symbols), along with those of SrTiO$_{3-\delta}$ (black symbols) and Nb:SrTiO$_3$ (red symbols) (refs.[44,64–66]). The solid lines represent the characteristics of a phenomenological model[67], $\mu \propto n^{-\frac{5}{6}}$, that fits the experimental data very well.

**Fig. 2 | Superconducting domes of SrTiO$_{3-\delta}$, La:SrTiO$_3$, and Nb:SrTiO$_3$. a,** Resistivity of our SrTi$_{1-x}$Nb$_x$O$_3$ single crystals (same samples studied in Fig. 1a) below 1 K for $0.0002 \le x \le 0.005$ (upper panel) and $0.007 \le x \le 0.05$ (lower panel). We defined $T_c$ as the mid-point of the resistance drop, as indicated by the three dotted lines for the $x$=0.007 data. The intersections of the lines give the onset and end of $T_c$, corresponding to the error bars shown in **b**. The $x$ = 0.0005 sample shows an indication of superconductivity below approximately 0.2 K, but zero resistivity was not achieved at the lowest measured temperature. The $T_c$ value increases with increasing $x$ up to $x$ = 0.005 and decreases for $x \ge 0.007$. **b,** Superconducting domes of Nb:SrTiO$_3$ obtained in this work (closed blue circles). The carrier density $n$ is the value at 5 K. The domes of our other works for La:SrTiO$_3$ (closed red squares[4] and triangles[42], with the three unpublished data points shown in Supplementary Figure 1) and La:SrTi($^{16}$O$_{0.4}$$^{18}$O$_{0.6}$)$_3$ (open red squares[4]) are also plotted. The dashed lines serve as a guide for the eyes. For comparison, other data from the literature are included for Nb:SrTiO$_3$ (closed green circles[23]), and SrTiO$_{3-\delta}$ (closed black triangles[49] and squares[12]).

**Fig. 3 | Resistance anomaly as evidence of existence of polar metal. a**, Resistivity values of our $Sr_{0.985}Ca_{0.015}Ti_{1-x}Nb_xO_3$ single crystals with $x$ = 0.0005, 0.001, 0.002, 0.005 and 0.015 plotted versus $T$. **b**, Curie temperature $T_C^{FE}$ ~25 K of the ferroelectric transition for undoped $Sr_{0.985}Ca_{0.015}TiO_3$ (open red circle) and temperatures $T_K$ at which the resistivity reaches a minimum for the doped samples (closed red triangles) plotted versus the carrier density $n$ at $T_K$ (or at 5 K for $T_K$ = 0). The values of $n$ at $T_K$ were obtained via interpolation (see the main text). The solid line represents the least-squares fit of $T_K$. This line is quite close to $T_C^{FE}$ at $n$ = 0. The inset shows the temperature dependence of the relative permittivity $\varepsilon$, where the peak corresponds to a $T_C^{FE}$ of ~25 K. This value is almost equal to that calculated using $T_C^{FE} = A(y - 0.0018)^{1/2}$ (where $A$ = 298 K and $y$ = 0.015) (ref.[32]). **c**, Resistivity of our $Sr_{0.95}Ba_{0.05}Ti_{1-x}Nb_xO_3$ single crystals with $x$ = 0.00025, 0.0003, 0.0005, 0.002, 0.0035, 0.005, 0.007, 0.015 and 0.02. **d**, $T_C^{FE}$~50 K for undoped $Sr_{0.95}Ba_{0.05}TiO_3$ (open blue circle), which corresponds to the peak of $\varepsilon$ (see the inset), and $T_K$ values for the doped samples (closed blue triangles) plotted versus $n$ at $T_K$. The definition of the error bars is provided in the main text. The solid line represents the least-squares fit for $T_K$.

**Fig. 4 | Superconductivity of the polar metals Nb:$Sr_{0.985}Ca_{0.015}TiO_3$ and Nb:$Sr_{0.95}Ba_{0.05}TiO_3$. a,** Resistivity of the $Sr_{0.985}Ca_{0.015}Ti_{1-x}Nb_xO_3$ single crystals (the same samples that were studied in Fig. 3**a**) below 1 K for 0.0005 ≤ $x$ ≤ 0.015. **b,** Resistivity of the $Sr_{0.95}Ba_{0.05}Ti_{1-x}Nb_xO_3$ single crystals (the same samples that were studied in Fig. 3**c**) below 1 K for 0.00025 ≤ $x$ ≤ 0.02. **c,** The bottom panel depicts the superconducting domes of the polar metals Nb:$Sr_{0.985}Ca_{0.015}TiO_3$ (denoted by Nb:Ca) and Nb:$Sr_{0.95}Ba_{0.05}TiO_3$ (Nb:Ba), along with that of the nonpolar Nb:$SrTiO_3$ (Nb:) that was shown in Fig. 2**b**. The carrier density $n$ of this plot was measured based on the Hall effect at 5 K. The dashed lines were obtained by performing a smoothing spline interpolation,

with the lines becoming negative in some regions. However, this behaviour is an artefact of the smoothing process and is negligibly small. Therefore, it does not affect the discussion here. The top panel shows a replot of the $T_K$ results shown in Fig. 3**b** and Fig. 3**d** versus the logarithmic values of $n$ at $T_K$. The dashed lines correspond to the solid lines shown in Fig. 3**b** and Fig. 3**d**. The systems are polar within the hatched regions. The solid lines represent the differences between the $T_c$ values of the polar metals and that of the nonpolar Nb:SrTiO$_3$. In both the top and bottom panels, the dash-dotted vertical lines correspond to $n^*$ for Nb:Sr$_{0.985}$Ca$_{0.015}$TiO$_3$ (light blue), and $n^*$ for Nb:Sr$_{0.95}$Ba$_{0.05}$TiO$_3$ (dark blue).

**Fig. 5 | $T_c$ comparison: resistivity, mutual inductance, and diamagnetism.** The top panel shows the superconducting transition observed via the AC mutual inductance (closed red circles) for Sr$_{0.95}$Ba$_{0.05}$Ti$_{0.995}$Nb$_{0.005}$O$_3$ (same sample in Fig. 3c) below 0.9 K in comparison with the resistive transition (closed blue circles) for the same sample. Although there must be a Joule heating effect due to the movement of the magnetic fluxes within the sample, the $T_c$ value that is defined as the onset of the mutual inductance drop (0.70 K) is almost equal to the corresponding resistive value (0.75 K). The bottom panel shows the DC volume magnetic susceptibility (Meissner effect) measured during warming after field cooling (FC) at 0.02, 0.05 and 0.1 mT for the same sample. The $T_c$ value (0.70 K) that is defined as the onset of diamagnetism nearly coincides with the zero resistivity temperature. (We have subtracted the background contributions here. See the Methods for full details.)

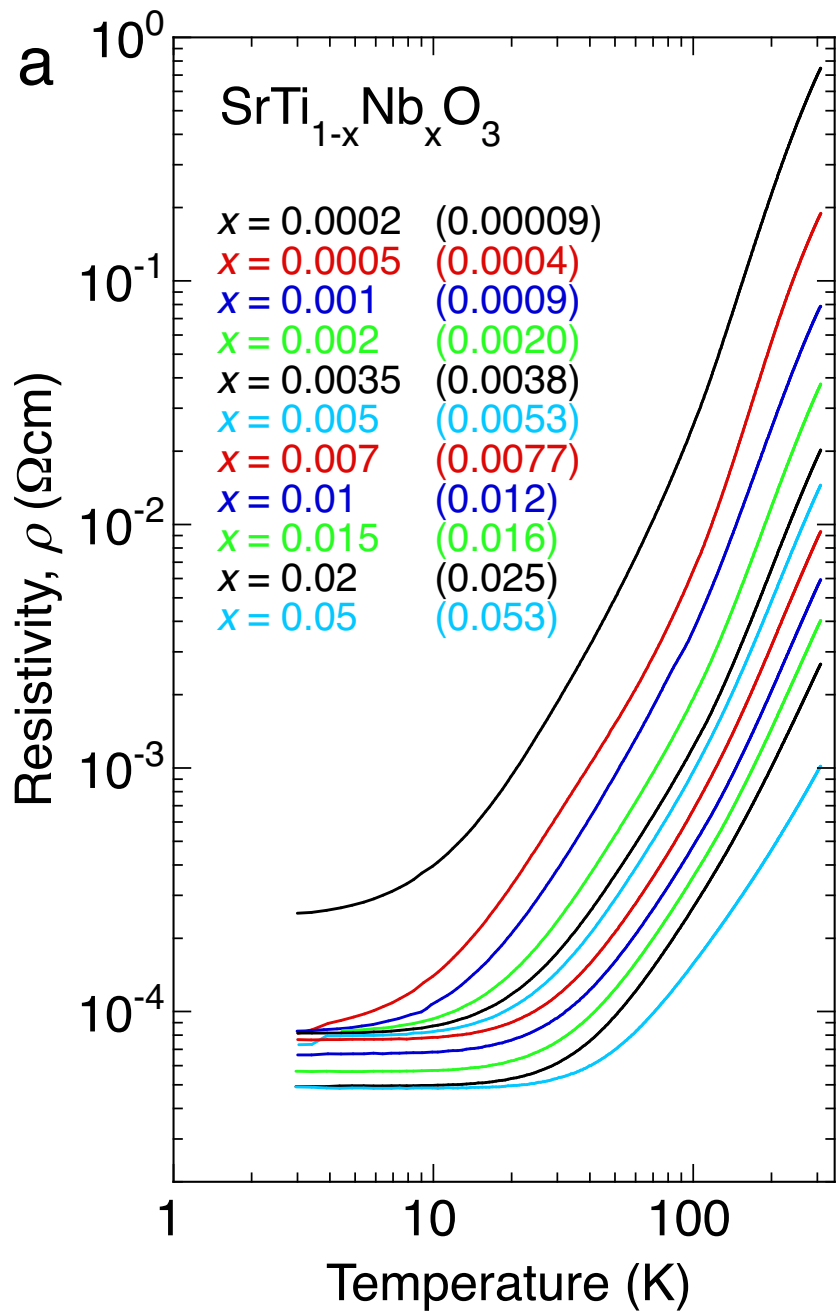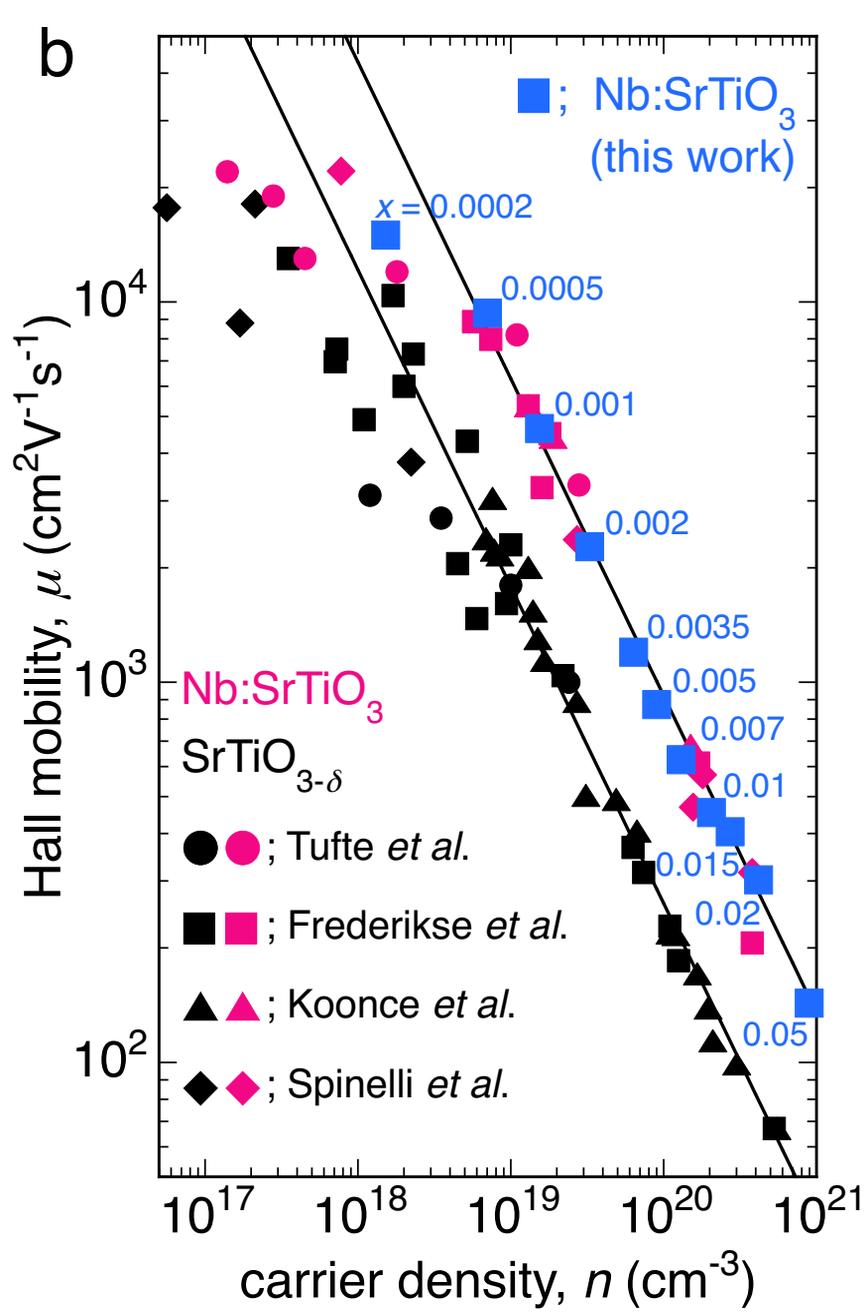

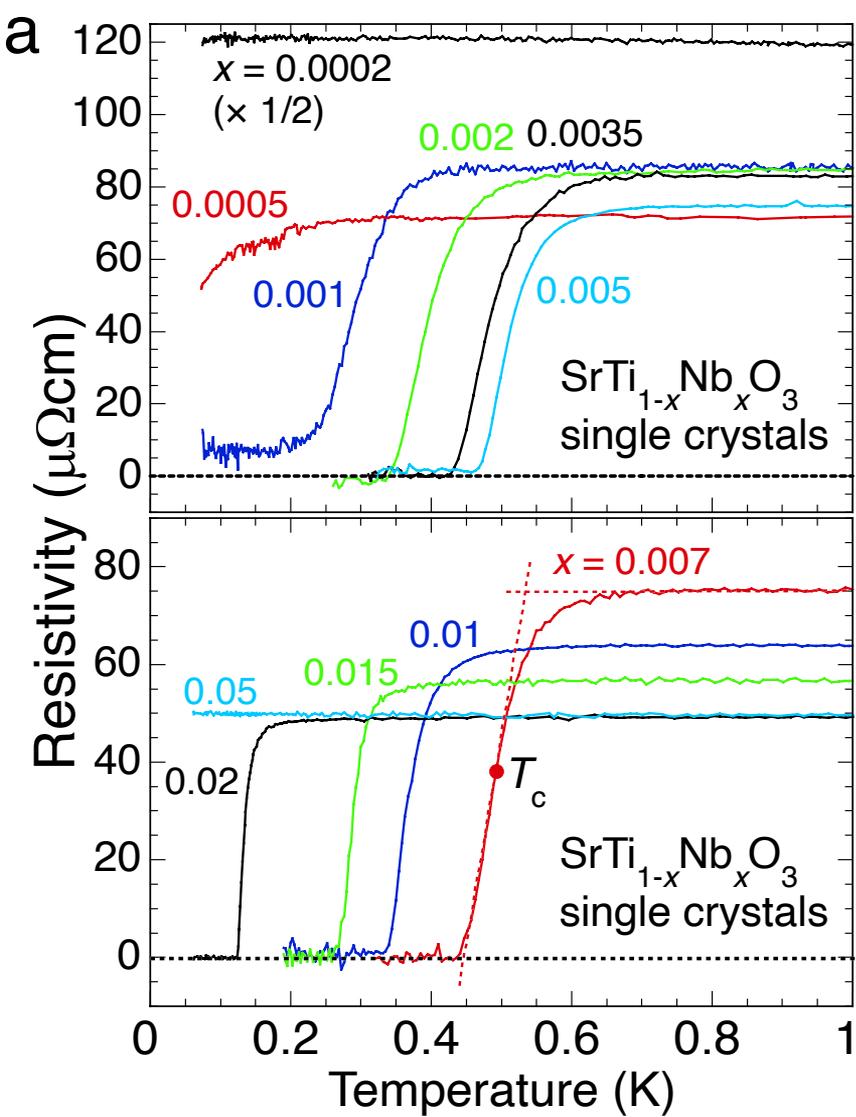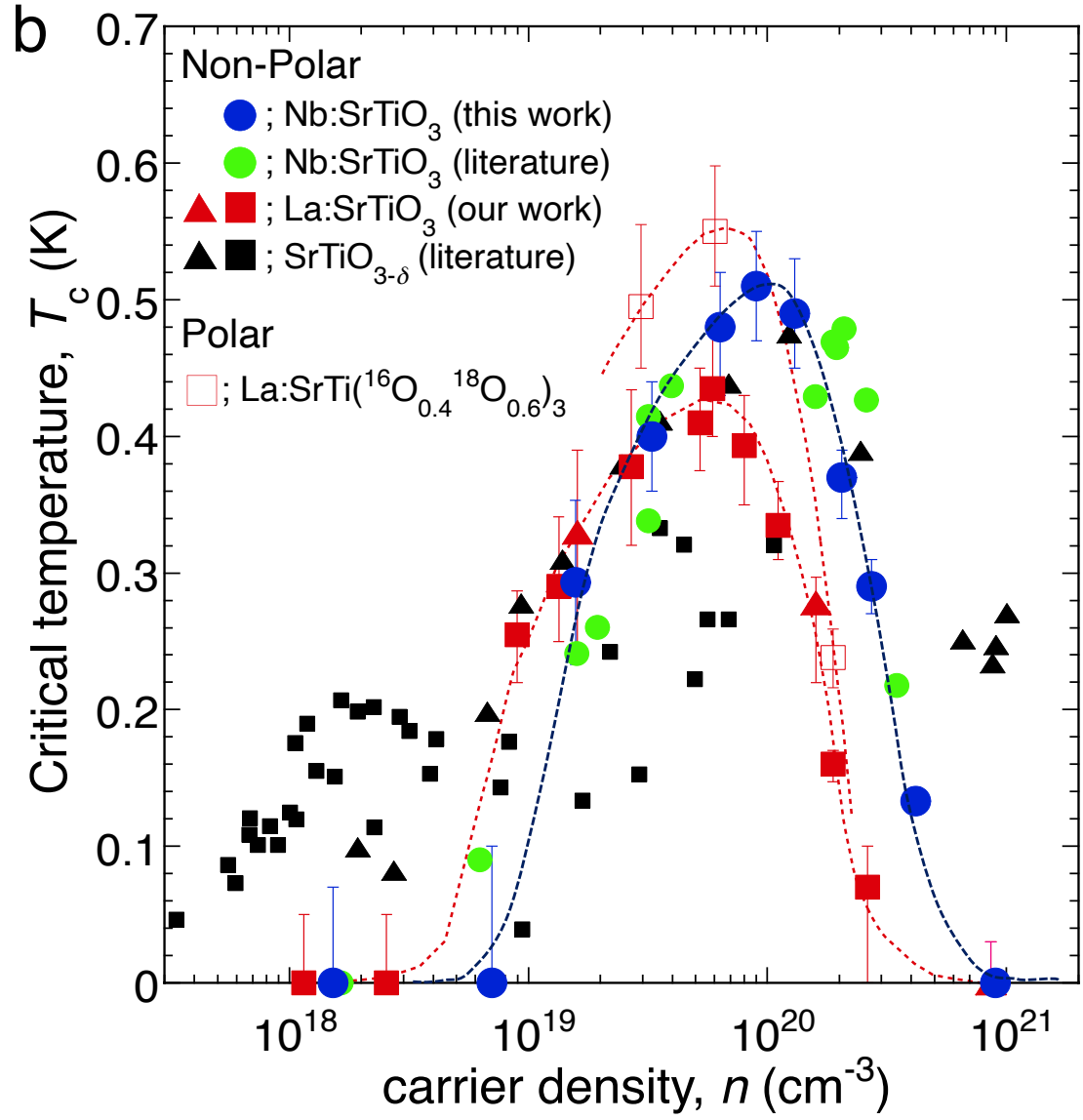

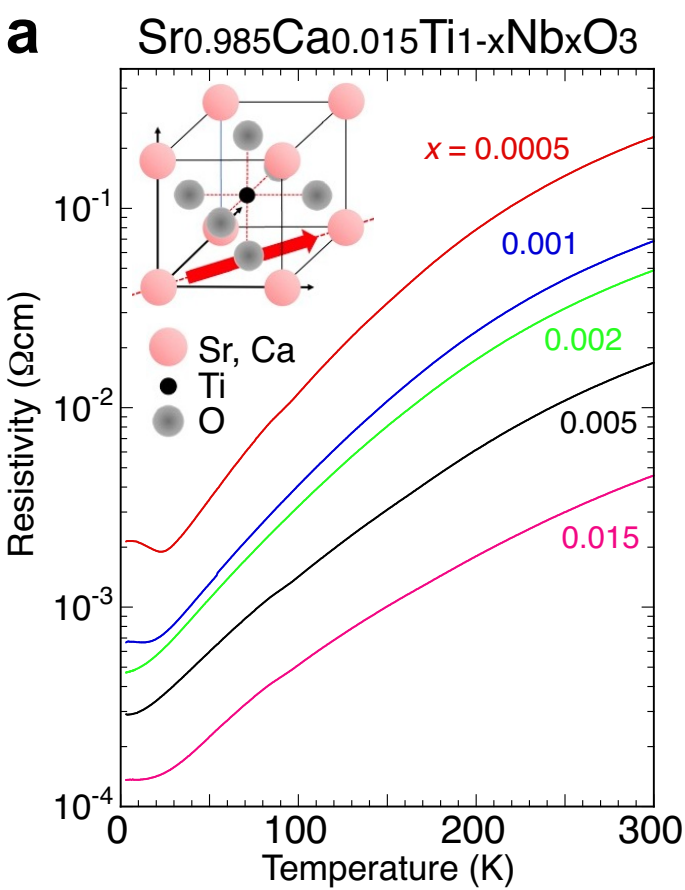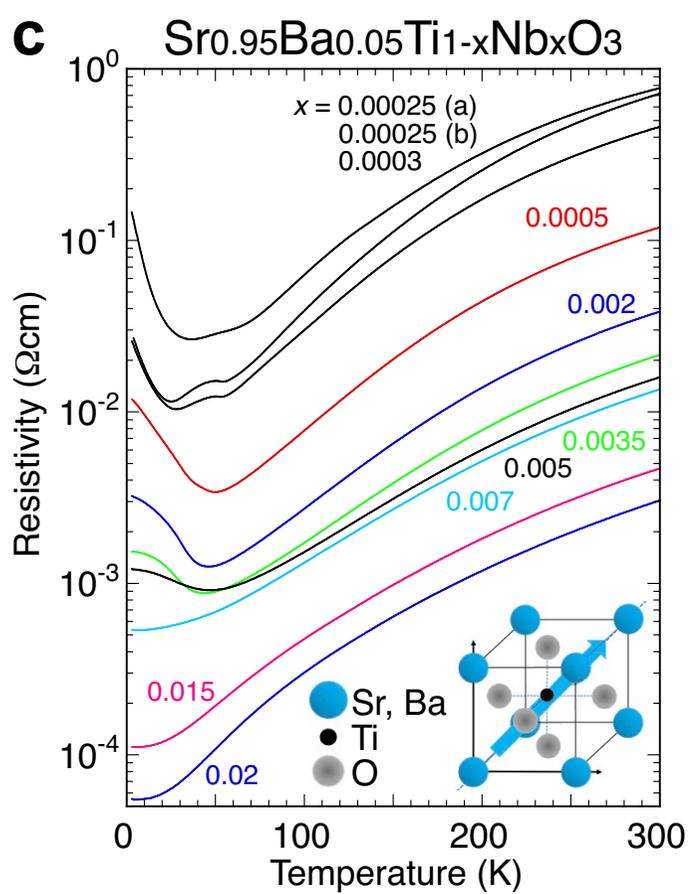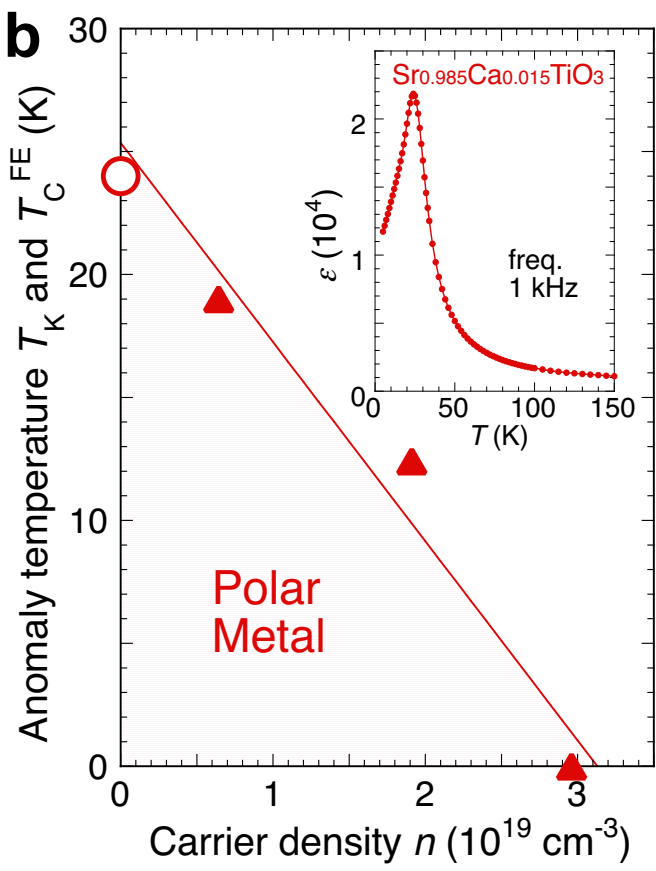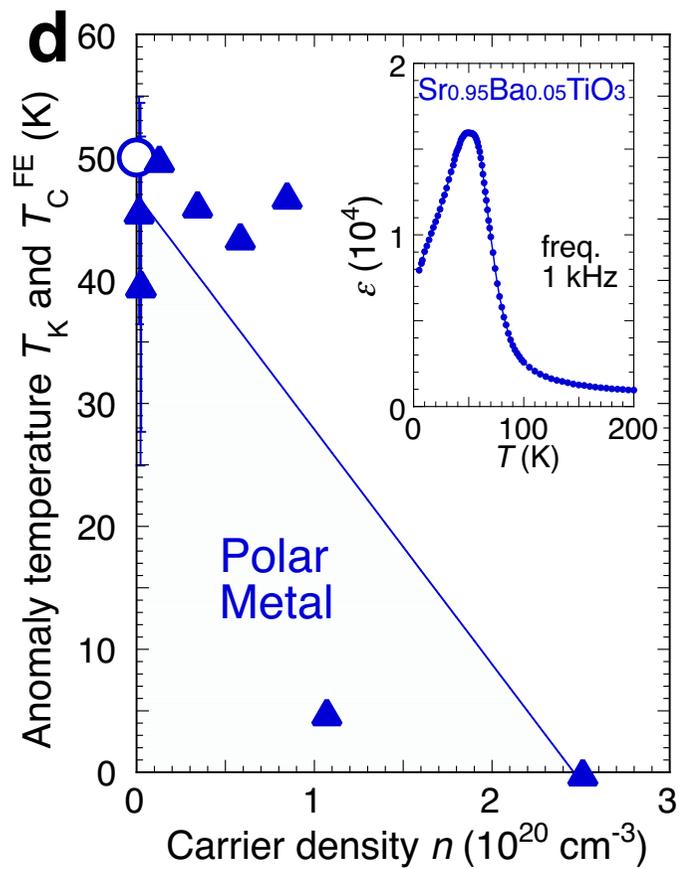

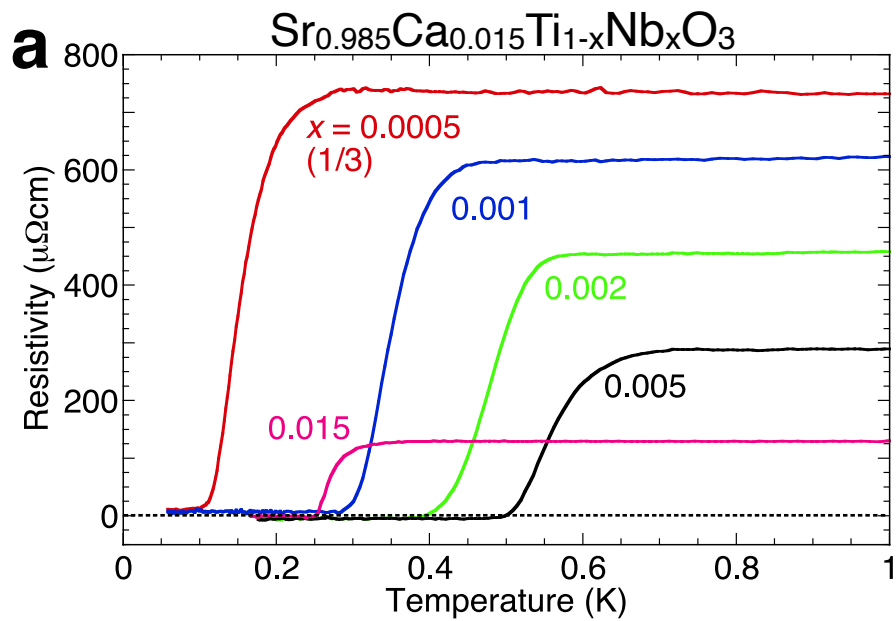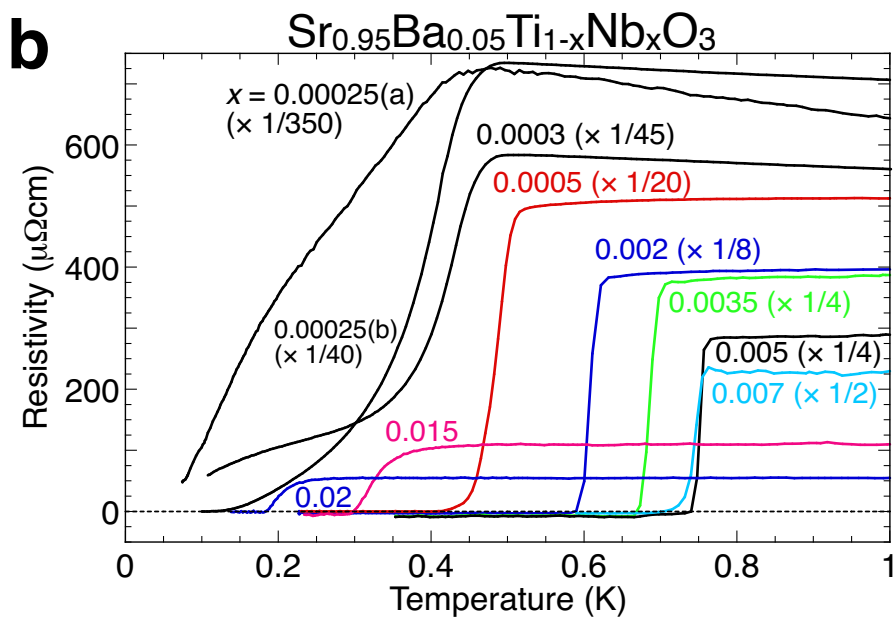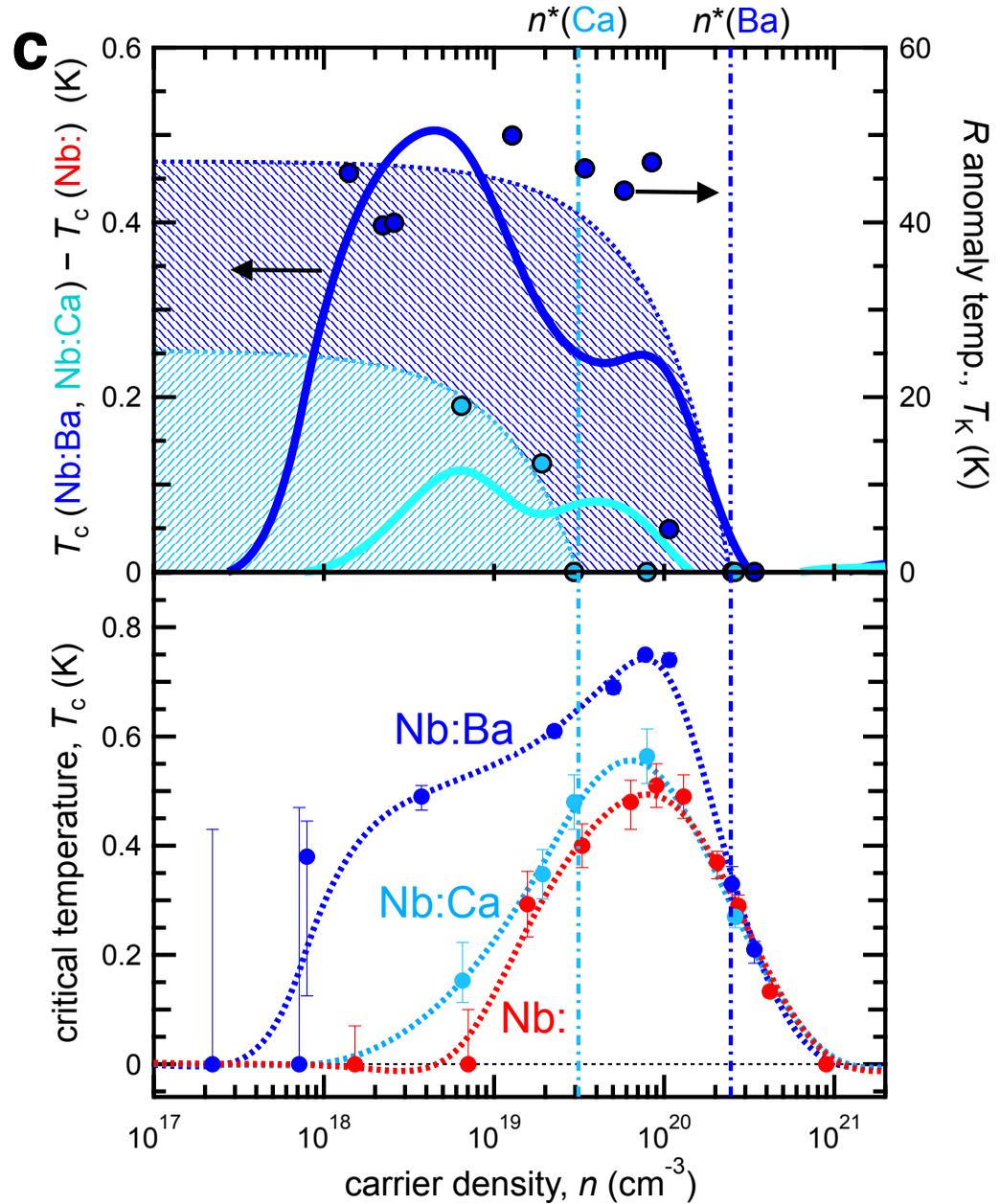

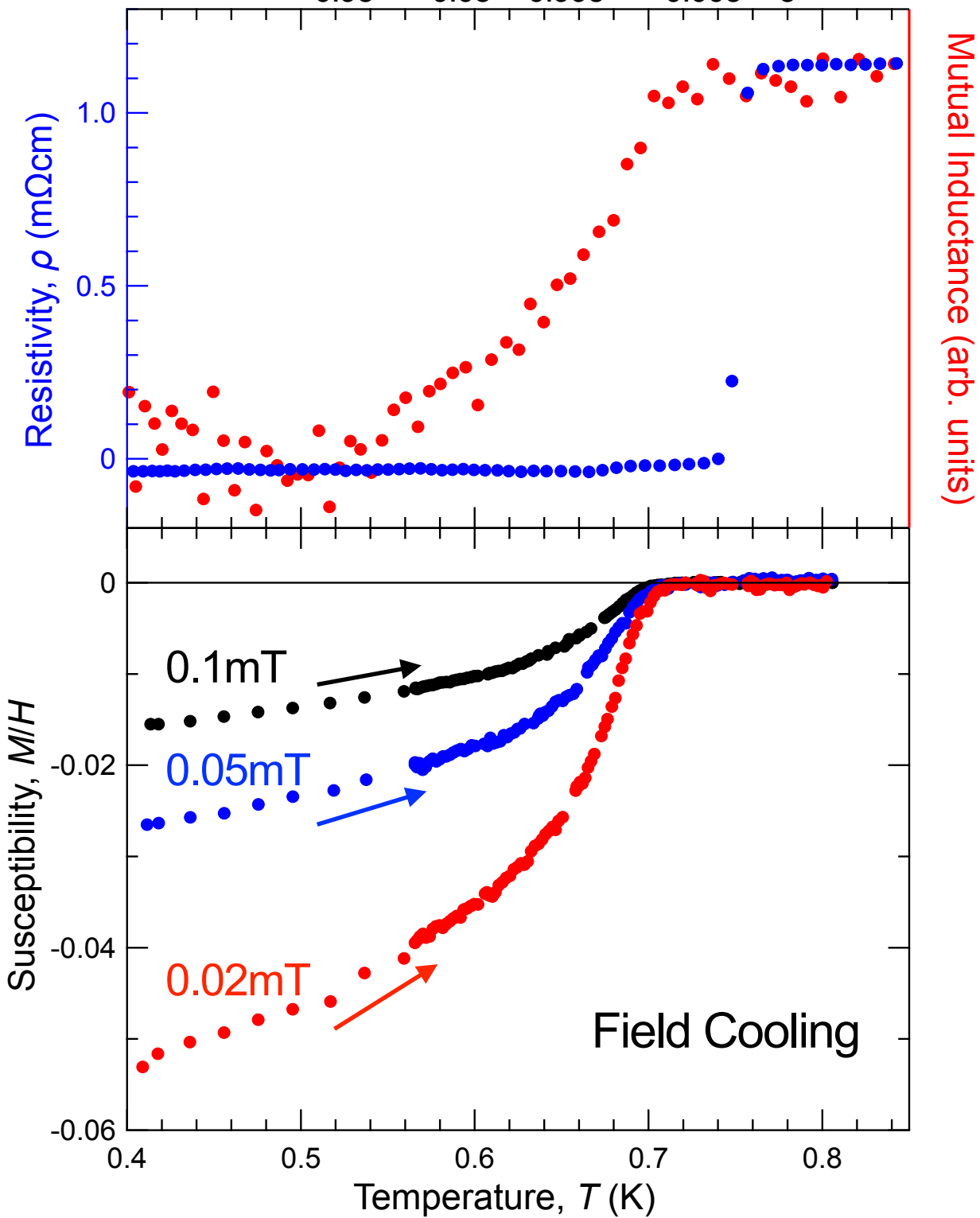

**Supplementary Information**

**Superconductivity enhancement in polar metal regions of  
$Sr_{0.95}Ba_{0.05}TiO_3$ and $Sr_{0.985}Ca_{0.015}TiO_3$  
revealed by systematic Nb doping**

Yasuhide Tomioka, Naoki Shirakawa, and Isao H. Inoue

National Institute of Advanced Industrial Science and Technology (AIST),  
Tsukuba 305-8565, Japan

**Supplementary Note 1: Update of our La:SrTiO$_3$ single crystals.**

In the previous paper, we reported the superconductivity of our Sr$_{1-x}$La$_x$TiO$_3$ single crystals[1]. La$^{3+}$ is a carrier dopant as Nb$^{5+}$ in SrTi$_{1-x}$Nb$_x$O$_3$ (denoted by Nb:SrTiO$_3$). Therefore, Sr$_{1-x}$La$_x$TiO$_3$ is referred to as La:SrTiO$_3$ in the main text. In this Supplementary Information, we follow the notation if it is not necessary to express it explicitly as SrTi$_{1-x}$Nb$_x$O$_3$ or Sr$_{1-x}$La$_x$TiO$_3$.

After the previous work, we continued preparing the La:SrTiO$_3$ samples, especially, in the dilute carrier-density regions ($x$ = 0.0001 and 0.0002) and the high carrier-density region ($x$ = 0.015). The results are shown in Supplementary Fig. 1. There is no significant upturn in the resistivity of La:SrTiO$_3$ at low temperatures. We think this is because La:SrTiO$_3$ is nonpolar as Nb:SrTiO$_3$.

In La:SrTiO$_3$, we could not observe symptoms of the superconductivity down to 50 mK for the $x$ = 0.0001 sample. The $x$ = 0.0002 sample showed an onset of the resistance drop without zero resistance. These results corroborate that there is no shoulder in the low carrier-density region on the superconducting dome of La:SrTiO$_3$ (see Fig. 2**b** in the main text). We could not observe the zero resistance for the $x$ = 0.015 sample, though there is a symptom of superconductivity. Like all the doped SrTiO$_3$ families, superconductivity in La:SrTiO$_3$ disappears at high carrier-density regions.

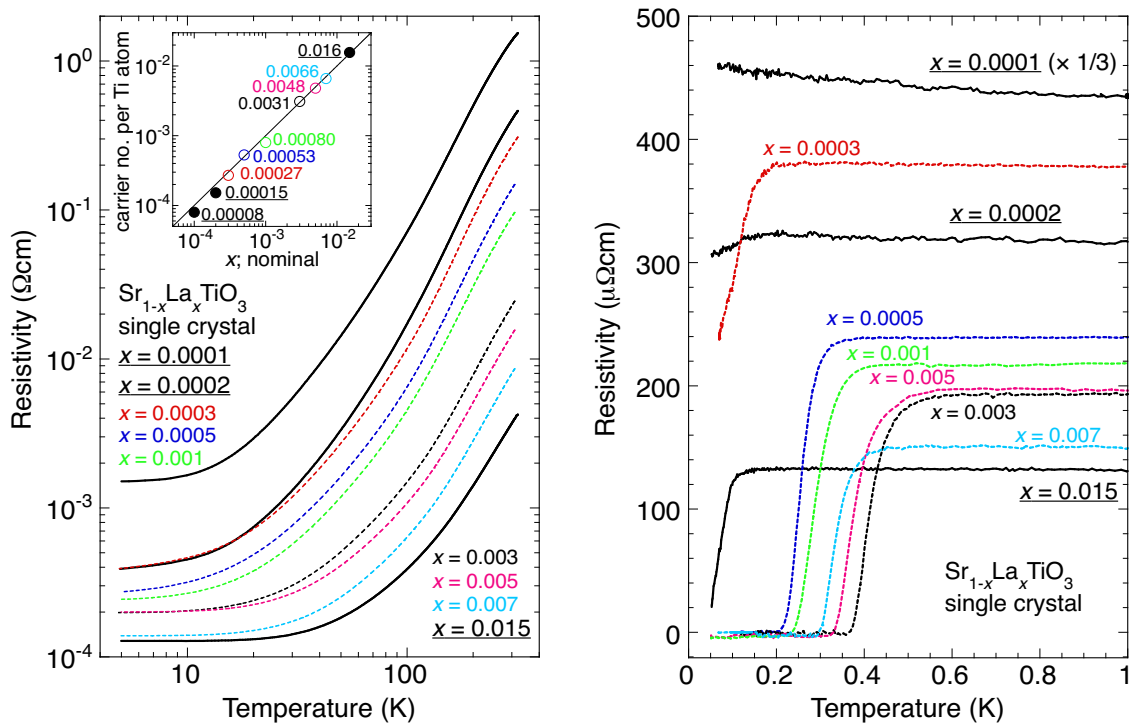

**Supplementary Figure 1 | Temperature dependence of the resistivity for La:SrTiO$_3$. Left**, Resistivity from room temperature down to 5 K of the La:SrTiO$_3$ single crystals with nominal $x$ values of 0.0001, 0.0002, 0.0003, 0.0005, 0.001, 0.003, 0.005, 0.007, and 0.015. The number of electrons per Ti site determined using the Hall effect measurements was almost identical to the nominal $x$ values (Inset). **Right**, Resistivity below 1 K shows superconductivity.

**Supplementary Note 2: Sample details for the Nb:SrTiO$_3$ single crystals.**

Sample details and measured numbers obtained by our study for the Nb:SrTiO$_3$ single crystals are summarised in Supplementary Table 1. The carrier density $n$ at 5 K was estimated as described in Supplementary Note 3. The Hall mobility $\mu$ at 5 K was calculated by $\mu = (ne\rho)^{-1}$ using the values of $n$ and $\rho$ at 5 K.

**Supplementary Table 1 | Sample details for the Nb:SrTiO$_3$ single crystals.** Carrier density $n$ (per formula unit and per volume in cm$^3$) determined from the Hall resistivity at 5K, the resistivity $\rho$ at 300 K and 5 K, the Hall mobility $\mu$ at 5 K, and the critical temperature for superconductivity $T_c$ obtained by our measurements. See main text for the definition of error bars.

| x (nominal) | $n$ (5 K) (per f.u.) | $n$ (5 K) (cm$^{-3}$) | $\rho$ (300 K) ($\Omega$cm) | $\rho$ (5 K) ($\Omega$cm) | $\mu$ (5 K) (cm$^2$V$^{-1}$s$^{-1}$) | $T_c$ (K) |
|---|---|---|---|---|---|---|
| 0.0002 | (9.0±0.2) × 10$^{-5}$ | (1.52±0.03) × 10$^{18}$ | (6.97±0.26) × 10$^{-1}$ | (2.74±0.1) × 10$^{-4}$ | (1.50±0.03) × 10$^{4}$ | (< 0.1) |
| 0.0005 | (4.2±0.1) × 10$^{-4}$ | (7.03±0.07) × 10$^{18}$ | (1.76±0.07) × 10$^{-1}$ | (9.49±0.4) × 10$^{-5}$ | (9.36±0.09) × 10$^{3}$ | (< 0.1) |
| 0.001 | (9.3±0.2) × 10$^{-4}$ | (1.56±0.03) × 10$^{19}$ | (7.36±0.3) × 10$^{-2}$ | (8.59±0.3) × 10$^{-5}$ | (4.66±0.09) × 10$^{3}$ | 0.29±0.06 |
| 0.002 | (2.0±0.02) × 10$^{-3}$ | (3.28±0.03) × 10$^{19}$ | (3.51±0.1) × 10$^{-2}$ | (8.38±0.3) × 10$^{-5}$ | (2.27±0.02) × 10$^{3}$ | 0.40±0.06 |
| 0.0035 | (3.8±0.04) × 10$^{-3}$ | (6.34±0.06) × 10$^{19}$ | (1.88±0.07) × 10$^{-2}$ | (8.17±0.3) × 10$^{-5}$ | (1.21±0.01) × 10$^{3}$ | $0.48^{+0.05}_{-0.04}$ |
| 0.005 | (5.3±0.05) × 10$^{-3}$ | (8.97±0.09) × 10$^{19}$ | (1.34±0.05) × 10$^{-2}$ | (7.97±0.3) × 10$^{-5}$ | (8.73±0.09) × 10$^{2}$ | 0.51±0.04 |
| 0.007 | (7.7±0.06) × 10$^{-3}$ | (1.30±0.01) × 10$^{20}$ | (8.72±0.3) × 10$^{-3}$ | (7.69±0.3) × 10$^{-5}$ | (6.24±0.05) × 10$^{2}$ | 0.49±0.04 |
| 0.01 | (1.2±0.01) × 10$^{-2}$ | (2.05±0.02) × 10$^{20}$ | (5.54±0.2) × 10$^{-3}$ | (6.72±0.3) × 10$^{-5}$ | (4.53±0.05) × 10$^{2}$ | $0.37^{+0.02}_{-0.03}$ |
| 0.015 | (1.6±0.02) × 10$^{-2}$ | (2.72±0.03) × 10$^{20}$ | (3.77±0.1) × 10$^{-3}$ | (5.68±0.2) × 10$^{-5}$ | (4.04±0.04) × 10$^{2}$ | 0.29±0.02 |
| 0.02 | (2.5±0.02) × 10$^{-2}$ | (4.17±0.04) × 10$^{20}$ | (2.47±0.1) × 10$^{-3}$ | (4.96±0.2) × 10$^{-5}$ | (3.02±0.03) × 10$^{2}$ | 0.13±0.01 |
| 0.05 | (5.3±0.05) × 10$^{-2}$ | (8.96±0.09) × 10$^{20}$ | (9.63±0.4) × 10$^{-4}$ | (4.86±0.2) × 10$^{-5}$ | (1.43±0.02) × 10$^{2}$ | (< 0.1) |

**Supplementary Note 3: Carrier-density estimation for the Nb:SrTiO$_3$ single crystals.**

The Hall effect for our Nb:SrTiO$_3$ single crystals was almost independent of temperature. We plotted the Hall resistivity $\rho_H$ as a function of the magnetic field $\mu_0 H$ ($\mu_0$ is the permeability of the vacuum) at 5 K in Supplementary Fig. 2. We tried to fit the linear relationship $\rho_H = \mu_0 H (en)^{-1}$ to the experimental data using the least squares method and deduced the carrier density $n$, where $e$ is the elementary charge. As seen in Supplementary Fig. 2, the fitting is reasonably good, with the error bars of $n$ being the standard deviation of the least squares fit. The number of electrons per formula unit (f.u.) is deduced from $n$, almost equivalent to the nominal number of $x$, so we used this $x$ value throughout the main text.

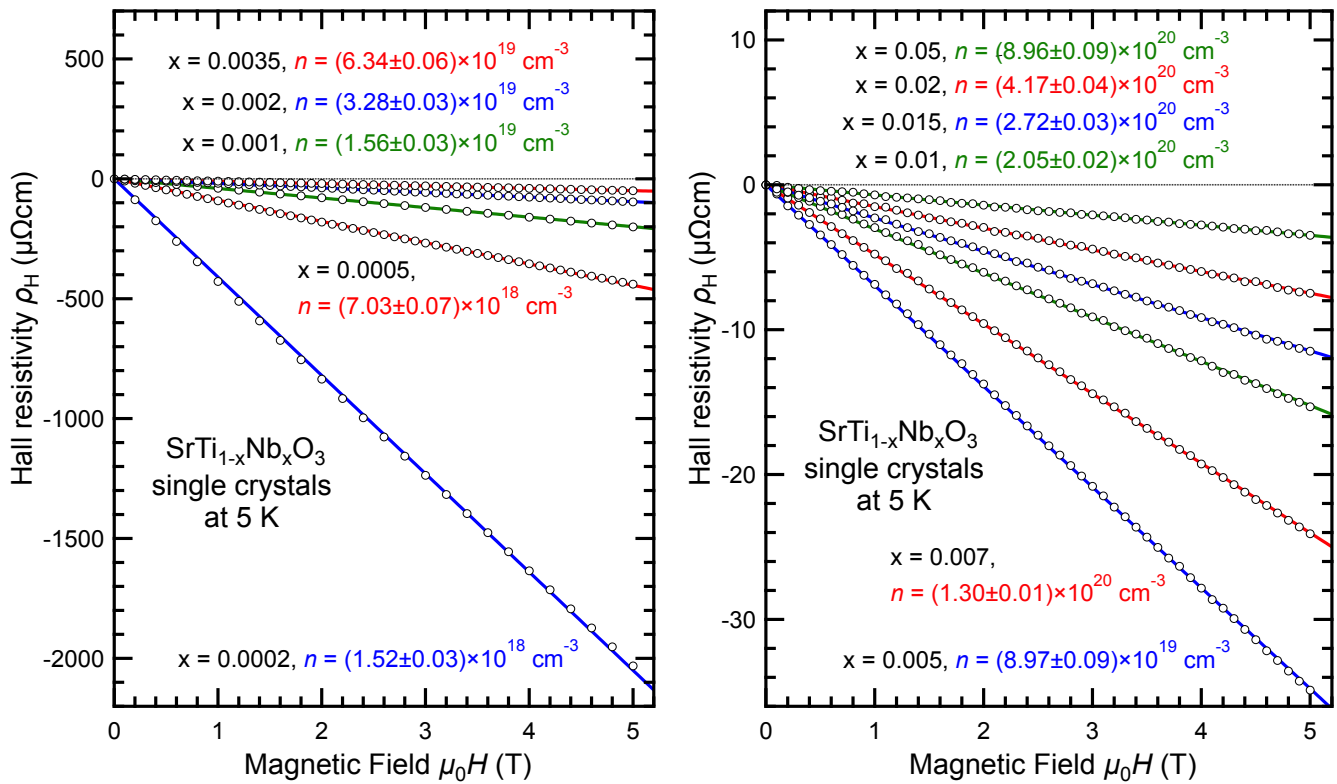

**Supplementary Figure 2 | Hall effect at 5 K for Nb:SrTiO$_3$ single crystals.** Hall resistivity $\rho_H$ (open circles) versus magnetic field $\mu_0 H$ shows linear relationships. The carrier density $n$ was deduced from this plot by $\rho_H = \mu_0 H (en)^{-1}$ (thick straight lines) using the least squares method. The error bars for $n$ correspond to the standard deviation.

**Supplementary Note 4: Hall mobility of the Nb:SrTiO₃ single crystals.**

Supplementary Figure 3 shows the logarithmic plot of the Hall mobility $\mu$ at 5 K of our Nb:SrTiO₃ single crystals versus the carrier density $n$ at 5 K. In the standard metal, $\mu$ is proportional to $n^{-1}$, while Behnia proposed $\mu \propto an^{-5/6}$ in his model[2]. We tried to fit $\mu \propto an^{-b}$ to our experimental data by the least squares method, and the results are shown in Supplementary Fig. 3. The best fit is given by $b = 0.85 \pm 0.02$. This value of $b$ means that $b = 5/6$ falls within the error bar, whereas b = 1 does not.

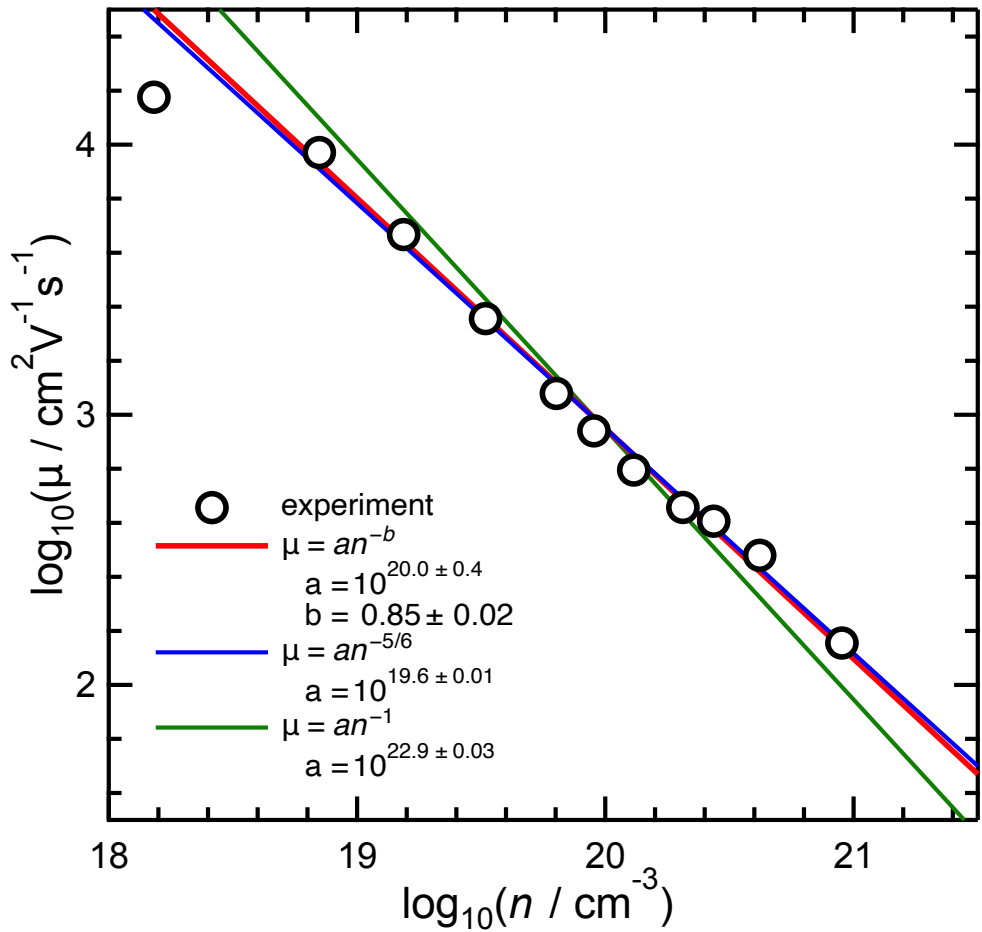

**Supplementary Figure 3 | Logarithmic plot of the Hall mobility $\mu$ vs the carrier density $n$ of our Nb:SrTiO₃ single crystals at 5 K.** The experimental values (open circles) are fitted by the $\mu \propto an^{-b}$ relationship (solid red line) by the least squares method (the error bars correspond to the standard deviations). The solid blue line corresponds to $b = 5/6$ (ref.[2]), while the solid green line corresponds to the behaviour seen in ordinary metals.

**Supplementary Note 5: Sample details for the polar Nb:Sr$_{0.985}$Ca$_{0.015}$TiO$_3$ single crystals.**

Like the nonpolar Nb:SrTiO$_3$, the Hall resistivity $\rho_H$ at 5 K is linear to the magnetic field $\mu_0 H$, and the carrier density *n* was evaluated by $\rho_H = \mu_0 H(en)^{-1}$ using the least squares method, the same as shown in Supplementary Fig. 2. The number of electrons per formula unit (f.u.) is almost equivalent to the nominal number of *x*, so we used this *x* value throughout the main text. The resistivity as a function of temperature shows an upturn for low carrier-density samples, and the temperature for the anomaly (i.e., the temperature which gives local minimum of the resistivity) is referred to as $T_K$. Sample details and measured numbers obtained by our study for the Nb:Sr$_{0.985}$Ca$_{0.015}$TiO$_3$ single crystals are summarised in Supplementary Table 2.

Stoichiometry of cations was checked using the inductively coupled plasma atomic emission spectroscopy (ICP-AES) for Sr$_{0.985}$Ca$_{0.015}$TiO$_3$ and Nb:Sr$_{0.985}$Ca$_{0.015}$TiO$_3$ (Nb 0.5%) single crystals. The results are listed in Table 3.

**Supplementary Table 2 | Sample details for the polar Nb:Sr$_{0.985}$Ca$_{0.015}$TiO$_3$ single crystals.** Carrier density *n* (per formula unit and volume in cm$^3$) determined from the Hall resistivity at 5 K, the resistivity $\rho$ at 300 K and 5 K, the resistance anomaly temperature $T_K$, and the critical temperature for superconductivity $T_c$ obtained by our measurements.

| *x* (nominal) | *n* (5 K) (per f.u.) | *n* (5 K) (cm$^{-3}$) | $\rho$(300 K) ($\Omega$cm) | $\rho$(5 K) ($\Omega$cm) | $T_K$ (K) | $T_c$ (K) |
|---|---|---|---|---|---|---|
| 0.0005 | (3.9±0.3) × 10$^{-4}$ | (6.52±0.01) × 10$^{18}$ | (2.27±0.09) × 10$^{-1}$ | (2.15±0.08) × 10$^{-3}$ | 19 | $0.15^{+0.07}_{-0.04}$ |
| 0.001 | (1.1±0.1) × 10$^{-3}$ | (1.93±0.001) × 10$^{19}$ | (6.85±0.3) × 10$^{-2}$ | (6.70±0.3) × 10$^{-4}$ | 12.4 | 0.35±0.045 |
| 0.002 | (1.8±0.1) × 10$^{-3}$ | (2.96±0.002) × 10$^{19}$ | (4.88±0.2) × 10$^{-2}$ | (4.75±0.2) × 10$^{-4}$ | | 0.48±0.05 |
| 0.005 | (4.7±0.4) × 10$^{-3}$ | (7.92±0.004) × 10$^{19}$ | (1.68±0.06) × 10$^{-2}$ | (2.89±0.1) × 10$^{-4}$ | | 0.56±0.05 |
| 0.015 | (1.6±0.1) × 10$^{-2}$ | (2.61±0.04) × 10$^{20}$ | (4.57±0.2) × 10$^{-2}$ | (1.36±0.05) × 10$^{-4}$ | | $0.27^{+0.01}_{-0.02}$ |

**Supplementary Table 3 | Cation ratios for Sr$_{0.985}$Ca$_{0.015}$TiO$_3$ and Nb:Sr$_{0.985}$Ca$_{0.015}$TiO$_3$ (Nb 0.5%) single crystals.** The Ca/(Sr + Ca) and Nb/(Nb + Ti) ratios were determined from the ICP-AES analyses.

| Nb/(Ti + Nb) | | Ca/(Sr + Ca) | |
|---|---|---|---|
| (nominal) | (measured) | (nominal) | (measured) |
| 0 | 0 | 0.015 | (1.5±0.05) × 10$^{-2}$ |
| 0.005 | (4.9±0.05) × 10$^{-3}$ | 0.015 | (1.5±0.05) × 10$^{-2}$ |

**Supplementary Note 6: Sample details for the polar Nb:Sr$_{0.95}$Ba$_{0.05}$TiO$_3$ single crystals.**

Like nonpolar Nb:SrTiO$_3$ and polar Nb:Sr$_{0.985}$Ca$_{0.015}$TiO$_3$, $n$ is evaluated as shown in Supplementary Fig. 2. The number of electrons per formula unit (f.u.) is almost equivalent to the nominal number of $x$, so we used this $x$ value throughout the main text. $T_K$ is the temperature for the resistivity minimum. For samples with two local minima in their resistivity characteristics, the mid-point between the two temperatures that give these minima is defined as $T_K$, where the lower and upper ends of the error bar correspond to each of these local minima. Sample details and measured numbers obtained by our study for the Nb:Sr$_{0.95}$Ba$_{0.05}$TiO$_3$ single crystals are summarised in Supplementary Table 4.

Stoichiometry of cations was checked using the ICP-AES for Sr$_{0.95}$Ba$_{0.05}$TiO$_3$ and Nb:Sr$_{0.95}$Ba$_{0.05}$TiO$_3$ (Nb 0.2%) single crystals. The results are listed in Supplementary Table 5.

**Supplementary Table 4 | Sample details for the polar Nb:Sr$_{0.95}$Ba$_{0.05}$O$_3$ single crystals.**
Carrier density $n$ (per a formula unit and volume in cm$^3$) determined from the Hall resistivity at 5 K, the resistivity $\rho$ at 300 K and 5 K, the resistance anomaly temperature $T_K$, and the critical temperature for superconductivity $T_c$ obtained by our measurements.

| $x$ (nominal) | $n$(5 K) (per f.u.) | $n$(5 K) (cm$^{-3}$) | $\rho$(300 K) ($\Omega$cm) | $\rho$(5 K) ($\Omega$cm) | $T_K$ (K) | $T_c$ (K) |
|---|---|---|---|---|---|---|
| 0.00025(a) | $(1.3\pm0.1) \times 10^{-5}$ | $(2.22\pm0.001) \times 10^{17}$ | $(7.72\pm0.3) \times 10^{-1}$ | $(1.18\pm0.03) \times 10^{-1}$ | 45.7±9 | $0.00^{+0.43}_{-0.00}$ |
| 0.00025(b) | $(4.7\pm0.4) \times 10^{-5}$ | $(7.98\pm0.001) \times 10^{17}$ | $(7.17\pm0.3) \times 10^{-1}$ | $(2.51\pm0.09) \times 10^{-2}$ | 39.7±15 | $0.38^{+0.07}_{-0.26}$ |
| 0.0003 | $(4.3\pm0.3) \times 10^{-5}$ | $(7.18\pm0.001) \times 10^{17}$ | $(4.59\pm0.2) \times 10^{-1}$ | $(2.29\pm0.09) \times 10^{-2}$ | 39.7±12 | $0.00^{+0.47}_{-0.00}$ |
| 0.0005 | $(2.2\pm0.2) \times 10^{-4}$ | $(3.75\pm0.003) \times 10^{18}$ | $(1.19\pm0.04) \times 10^{-1}$ | $(1.13\pm0.04) \times 10^{-2}$ | 49.9 | $0.49^{+0.02}_{-0.03}$ |
| 0.002 | $(1.3\pm0.1) \times 10^{-3}$ | $(2.27\pm0.001) \times 10^{19}$ | $(3.84\pm0.1) \times 10^{-2}$ | $(3.16\pm0.1) \times 10^{-3}$ | 46.2 | $0.61^{+0.01}_{-0.01}$ |
| 0.0035 | $(3.0\pm0.2) \times 10^{-3}$ | $(5.04\pm0.001) \times 10^{19}$ | $(2.15\pm0.08) \times 10^{-2}$ | $(1.53\pm0.06) \times 10^{-3}$ | 43.6 | $0.69^{+0.01}_{-0.01}$ |
| 0.005 | $(4.6\pm0.4) \times 10^{-3}$ | $(7.79\pm0.003) \times 10^{19}$ | $(1.59\pm0.06) \times 10^{-2}$ | $(1.20\pm0.05) \times 10^{-3}$ | 46.9 | $0.75^{+0.01}_{-0.01}$ |
| 0.007 | $(6.4\pm0.5) \times 10^{-3}$ | $(1.07\pm0.002) \times 10^{20}$ | $(1.35\pm0.05) \times 10^{-2}$ | $(5.33\pm0.2) \times 10^{-4}$ | 4.90 | $0.74^{+0.01}_{-0.01}$ |
| 0.015 | $(1.5\pm0.1) \times 10^{-2}$ | $(2.51\pm0.004) \times 10^{20}$ | $(4.67\pm0.2) \times 10^{-3}$ | $(1.11\pm0.04) \times 10^{-4}$ |  | $0.33^{+0.03}_{-0.03}$ |
| 0.02 | $(2.0\pm0.2) \times 10^{-2}$ | $(3.40\pm0.01) \times 10^{20}$ | $(3.04\pm0.1) \times 10^{-3}$ | $(5.50\pm0.2) \times 10^{-5}$ |  | $0.21^{+0.01}_{-0.03}$ |

**Supplementary Table 5 | Cation ratios for Sr$_{0.95}$Ba$_{0.05}$TiO$_3$ and Nb:Sr$_{0.95}$Ba$_{0.05}$TiO$_3$ (Nb 0.2%) single crystals.** The Ba/(Sr + Ba) and Nb/(Nb + Ti) ratios were determined from the ICP-AES analyses.

| Nb/(Ti + Nb) | | Ba/(Sr + Ba) | |
|---|---|---|---|
| (nominal) | (measured) | (nominal) | (measured) |
| 0 | 0 | 0.05 | $(3.9\pm0.1) \times 10^{-2}$ |
| 0.002 | $(2.0\pm0.05) \times 10^{-3}$ | 0.05 | $(5.5\pm0.1) \times 10^{-2}$ |

**Supplementary Note 7: Temperature dependence of the carrier density for the polar Nb:Sr$_{0.95}$Ba$_{0.05}$TiO$_3$ and Nb:Sr$_{0.985}$Ca$_{0.015}$TiO$_3$ single crystals.**

The carrier density $n$ is plotted as a function of temperature $T$ for Nb:Sr$_{0.95}$Ba$_{0.05}$TiO$_3$ (Supplementary Fig. 4, Left) and Nb:Sr$_{0.985}$Ca$_{0.015}$TiO$_3$ (Supplementary Fig. 4, Right). In Nb:Sr$_{0.95}$Ba$_{0.05}$TiO$_3$, $n$ decreases at low $T$. The onsets of the carrier-density drop roughly correspond to the resistance anomaly temperatures $T_K$. In contrast, in Nb:Sr$_{0.985}$Ca$_{0.015}$TiO$_3$, the drop in $n$ is not apparent. To derive $n$ at $T_K$, for Fig. 3**b** in the main text, we performed a smoothing spline interpolation[3] (smoothing factor = 1) for the $n$ versus $T$ data in Supplementary Fig. 4 using Igor Pro v8.04 software (WaveMetrics, Inc., USA). To avoid complicating the figure, the smoothing spline curves are not shown in Supplementary Fig. 4 (data points are connected by straight lines).

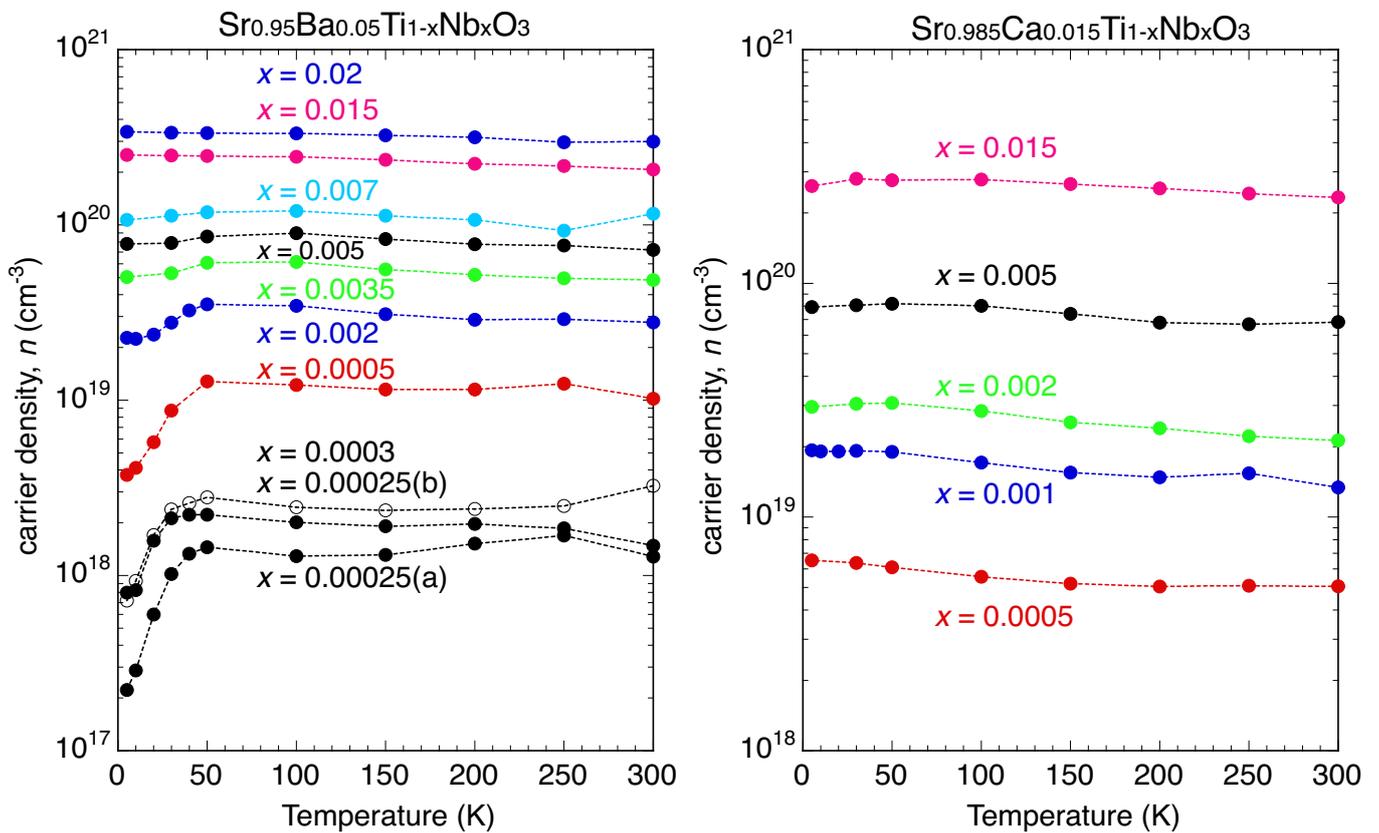

**Supplementary Figure 4 | Temperature dependence of the carrier density for polar Nb:Sr$_{0.95}$Ba$_{0.05}$TiO$_3$ and Nb:Sr$_{0.985}$Ca$_{0.015}$TiO$_3$ single crystals. Left,** in Nb:Sr$_{0.95}$Ba$_{0.05}$TiO$_3$, the carrier density $n$ (filled and open circles connected by dashed lines) decreases at low temperatures at which the resistivity shows anomaly. **Right,** in Nb:Sr$_{0.985}$Ca$_{0.015}$TiO$_3$, the temperature dependence is not apparent.

**Supplementary Note 8: $T_c$ of the polar Nb:Sr$_{0.95}$Ba$_{0.05}$TiO$_3$, Nb:Sr$_{0.985}$Ca$_{0.015}$TiO$_3$, and nonpolar Nb:SrTiO$_3$ plotted against $n$ at 5K and 50K.**

Since the carrier density $n$ of Nb:Sr$_{0.95}$Ba$_{0.05}$TiO$_3$ decreases below 50 K, as shown in Supplementary Fig. 4, a question arises as to which temperature $n$ should be used to plot the relationship between $T_c$ and $n$. We think it is physically more meaningful to use $n$ at a temperature as close to $T_c$ as possible, which is what we did in Fig. 4**c** in the main text. Here, however, we have plotted $T_c$ against $n$ at 50 K as a comparison (Supplementary Fig. 5). Although the persistence of $T_c$ in the low $n$ region appears to become smaller, the onset of $T_c$ is observed at much higher temperatures. Therefore, by connecting the onsets as indicated by the thick dashed line in Supplementary Fig. 5, it seems there is little difference between the two plots.

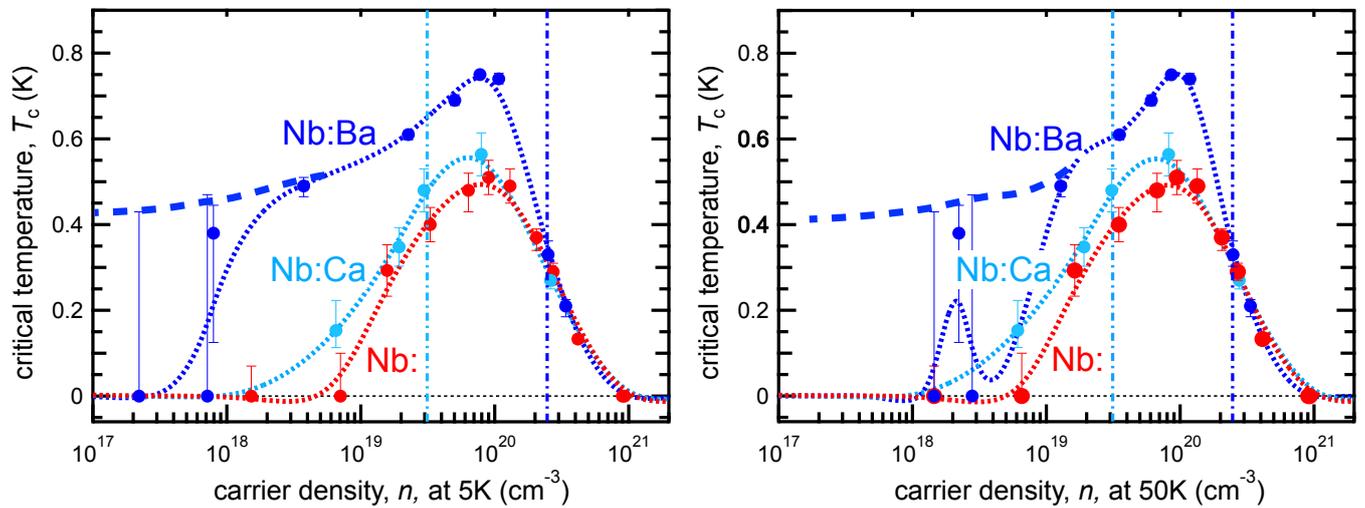

**Supplementary Figure 5 | Comparison of the $T_c$ vs $n$ plots for $n$ at 5 K and 50 K**. **Left**, same as the bottom panel of the Fig. 4**c** in the main text. For Nb:Sr$_{0.985}$Ca$_{0.015}$TiO$_3$ (denoted by Nb:Ca), Nb:Sr$_{0.95}$Ba$_{0.05}$TiO$_3$ (Nb:Ba), and Nb:SrTiO$_3$ (Nb:), $T_c$ vs $n$ at 5 K are shown. The thin dashed lines were obtained by performing a smoothing spline interpolation, and thick dashed line for Nb:Ba is connecting the onsets of $T_c$. The dash-dotted vertical lines correspond to $n^*$ for Nb:Sr$_{0.985}$Ca$_{0.015}$TiO$_3$ (light blue), and $n^*$ for Nb:Sr$_{0.95}$Ba$_{0.05}$TiO$_3$ (dark blue). **Right**, same as those in the **Left** figure, except for $n$ at 50 K.

**Supplementary Note 9: Ferroelectricity of non-doped Sr$_{0.985}$Ca$_{0.015}$TiO$_3$ and Sr$_{0.95}$Ba$_{0.05}$TiO$_3$ single crystals.**

The non-doped Sr$_{0.95}$Ba$_{0.05}$TiO$_3$ and Sr$_{0.985}$Ca$_{0.015}$TiO$_3$ single crystals show the ferroelectric phase transition at 50 K and 24 K, respectively, as summarised in Supplementary Table 6. The two ferroelectrics turn to polar metals with the resistance anomaly at low temperatures by doping a slight amount of Nb$^{5+}$ for Ti$^{4+}$, i.e., by electron doping. Then, by further Nb doping, the resistance anomaly disappears at the critical carrier density $n^*$ (Supplementary Fig. 6).

**Supplementary Table 6 |** Ferroelectric Curie temperature, the crystal symmetry of the ferroelectric phase (see below), the direction of polarisation, and carrier density $n^*$ above which the resistance anomaly disappears. The values of $n^*$ are deduced as shown in Supplementary Fig. 6.

|  | Curie temp. (K) | crystal symmetry | polarization direction | $n^*$ (cm$^{-3}$) |
|---|---|---|---|---|
| Sr$_{0.95}$Ba$_{0.05}$TiO$_3$ | 50 | $R$3m | [111] | (2.46±0.53) × 10$^{20}$ |
| Sr$_{0.985}$Ca$_{0.015}$TiO$_3$ | 24 | $P$4mm | [110] | (3.13±0.37) × 10$^{19}$ |

**Supplementary Note 10: Determining the boundary separating the polar-metal and nonpolar-metal phases for the Nb:Sr$_{0.95}$Ba$_{0.05}$TiO$_3$ and Nb:Sr$_{0.985}$Ca$_{0.015}$TiO$_3$ single crystals.**

We defined a specific temperature $T_K$ at which the resistivity reaches a minimum. For samples with two local minima in their resistivity characteristics, the mid-point between the two temperatures that give these minima is defined as $T_K$, where the lower and upper ends of the error bar correspond to each local minima. To derive $n$ at $T_K$, we performed a smoothing spline interpolation[3]. In Supplementary Fig. 6, the values of $T_K$ were plotted versus log($n$) at $T_K$. We used log($n$) at 5 K for $T_K = 0$ samples. The experimental data are fitted by the solid line $T_K = a(n - n^*)$ using the least squares method. The line corresponds to the boundary separating the polar-metal phase and the nonpolar-metal phase because the space symmetry is broken in the polar region, as shown in Supplementary Note 11, and Russel et al.[4] showed the evidence, i.e., the second harmonic generation signal showed a sharp increase at $T_K$. The values of $n^*$ are shown in Supplementary Fig. 6 and summarised in Supplementary Table 6.

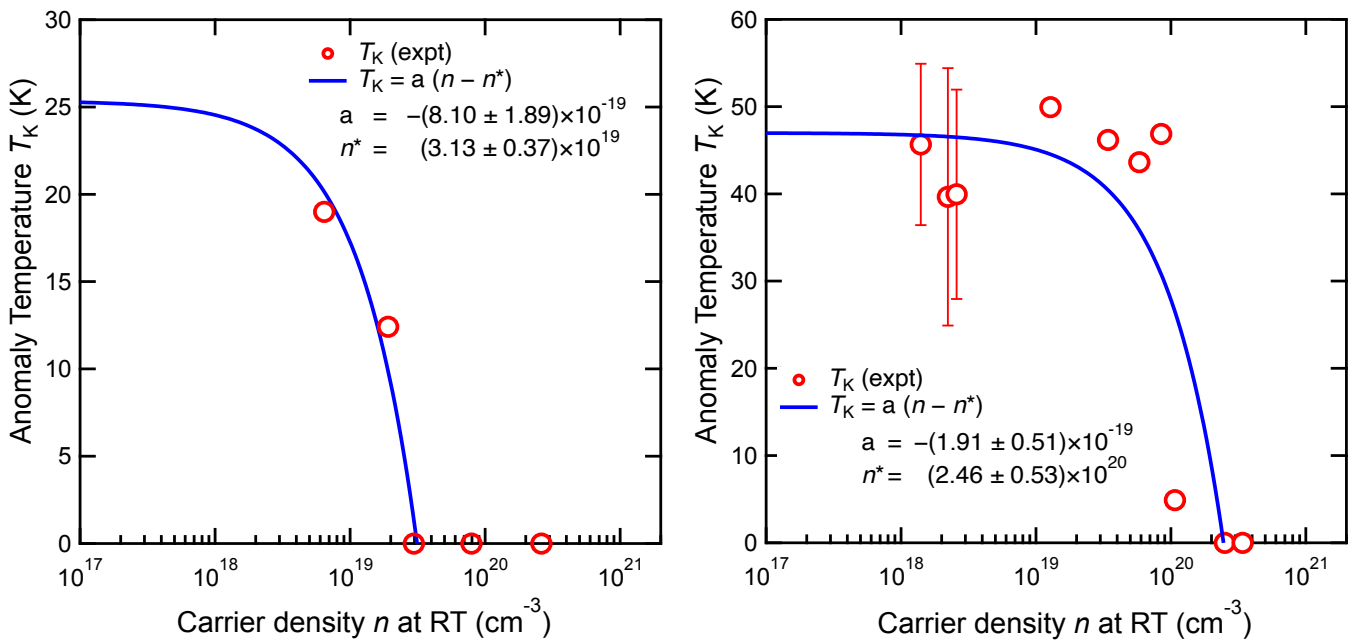

**Supplementary Figure 6 | $T_K$ as a function of the carrier density $n$ at $T_K$. Left**, a replot of the $T_K$ in Fig. 3**b** in the main text for Nb:Sr$_{0.985}$Ca$_{0.015}$TiO$_3$ versus the logarithmic values of $n$ at $T_K$. The solid blue line corresponds to the straight line in Fig. 3**b** in the main text obtained by fitting $T_K = a(n - n^*)$ to the experimental data (open red circles) using the least squares method. The error bars are the standard deviations. **Right**, a replot of the $T_K$ for Nb:Sr$_{0.95}$Ba$_{0.05}$TiO$_3$ in Fig. 3**d** in the main text versus the logarithmic values of $n$ at $T_K$. Same as the left figure, the solid blue line was obtained by fitting $T_K = a(n - n^*)$ to the experimental data (open red circles).

**Supplementary Note 11: Evidence of the space symmetry breaking in the metallic Nb:Sr$_{0.95}$Ba$_{0.05}$TiO$_3$ (Nb 0.2%) single crystal.**

Supplementary Table 7 shows the crystallographic data for our metallic Nb:Sr$_{0.95}$Ba$_{0.05}$TiO$_3$ (Nb 0.2%) single crystals. The powder X-ray diffraction pattern was analysed assuming a trigonal $R3m$ (hexagonal lattice) symmetry, and the results of the Rietveld refinement are summarised in Supplementary Table 8. The atomic coordination ($x$, $y$, $z$) of the B-site of the ABO$_3$ formula unit is (0, 0, 0.4957), indicating that the displacement of the B-site occurs along the [001] direction of the hexagonal lattice that is equivalent to the [111] direction of a pseudo-cubic lattice. These results indicate that, for Nb:Sr$_{0.95}$Ba$_{0.05}$TiO$_3$ (Nb 0.2%), the crystal structure at low temperatures is consistent with trigonal $R3m$, i.e., the centrosymmetricity is broken as in the ferroelectric matrix Sr$_{0.95}$Ba$_{0.05}$TiO$_3$.

**Supplementary Table 7 | Crystallographic data of the Nb:Sr$_{0.95}$Ba$_{0.05}$TiO$_3$ (Nb 0.2%) single crystal.** The lattice parameters can be expressed in the equivalent pseudo-cubic lattice as $a_R$ = 3.89826 Å and $α$ = 90.00°. The numbers in the parenthesis are the errors in the last digit, which is the standard uncertainties upon the Rietveld analysis.

| formula weight (g) | 186.061 |
|---|---|
| density (g/cm$^3$) | 5.2153 |
| crystal system | Trigonal (hexagonal lattice) |
| space group | $R3m$ |
| lattice parameters (Å) | $a = b = 5.513(6)$, $c = 6.751(8)$, $α = β = 90°$, $γ = 120°$ |

**Supplementary Table 8 | Refined atomic parameters at 10 K of the Nb:Sr$_{0.95}$Ba$_{0.05}$TiO$_3$ (Nb 0.2%) single crystal.** The Rietveld refinement was performed, converting the rhombohedral setting to a hexagonal one[5]. The reliability factor[6] of the weighted profile $R_{wp}$ = 14.8, and the patterns $R_p$ = 10.62. The $g$ and $B_{eq}$ are the occupancy and isotropic displacement parameters, respectively. The numbers in the parenthesis are the errors in the last digit, which is the standard uncertainties upon the Rietveld analysis.

|  | site | $x$ | $y$ | $z$ | $g$ | $B_{eq}$ (Å$^2$) |
|---|---|---|---|---|---|---|
| Sr$_{0.95}$Ba$_{0.05}$ | 3$a$ | 0 | 0 | 0 | 1 | 0.017 |
| Ti$_{0.998}$Nb$_{0.002}$ | 3$a$ | 0 | 0 | 0.4957 | 1 | 0.033 |
| O | 9$b$ | 0.499(2) | 0.500(2) | 0.014(3) | 1 | 0.006 |

**Supplementary Note 12: Variation of the polar Nb:Sr$_{0.95}$Ba$_{0.05}$TiO$_3$ samples in the dilute carrier-density region with the Nb contents of 0.025% and 0.2%.**

For the Nb:Sr$_{0.95}$Ba$_{0.05}$TiO$_3$ single crystals with the Nb content of 0.025%, the numbers of itinerant electrons at room temperature determined by the Hall effect measurements are both smaller than that of the nominal number of 0.00025. It was 0.00008 for the sample denoted by 25α and 0.00009 for 25β. The sample 25α corresponds to x=0.00025(a) and 25β does to x=0.00025(b) in the main text. A two-step resistance anomaly was seen in the samples 25β and 25γ, while the anomalies seem to be merged into one in the sample 25α (Supplementary Fig. 7**a**). As we decrease the temperature, the carrier density decreases from around 20 K. For the two-step $T_K$ samples, the onset temperature seems to correspond to the lower $T_K$ value (Supplementary Fig. 7**b**).

For the Nb:Sr$_{0.95}$Ba$_{0.05}$TiO$_3$ single crystals with the Nb content of 0.2%, the variations in $T_K$ was also observed (Supplementary Fig. 7**c**). In these samples, the numbers of the itinerant electrons at 100 K (0.002 for 200α and 0.0018 for 200β) are almost equal to the nominal number (0.002). All of them show the single resistance anomaly.

We have shown the data of 200α, 25α and 25β samples in the main text because we have not done the measurement of the superconductivity for 25γ, 200γ, and 200δ samples in the dilution refrigerater. For the 200β samples, althogh the residual resistivity is larger than that of the sample 200α, the superconducting transition temperature $T_c$ was almost comparable as shown in Supplementary Fig. 7**f**.

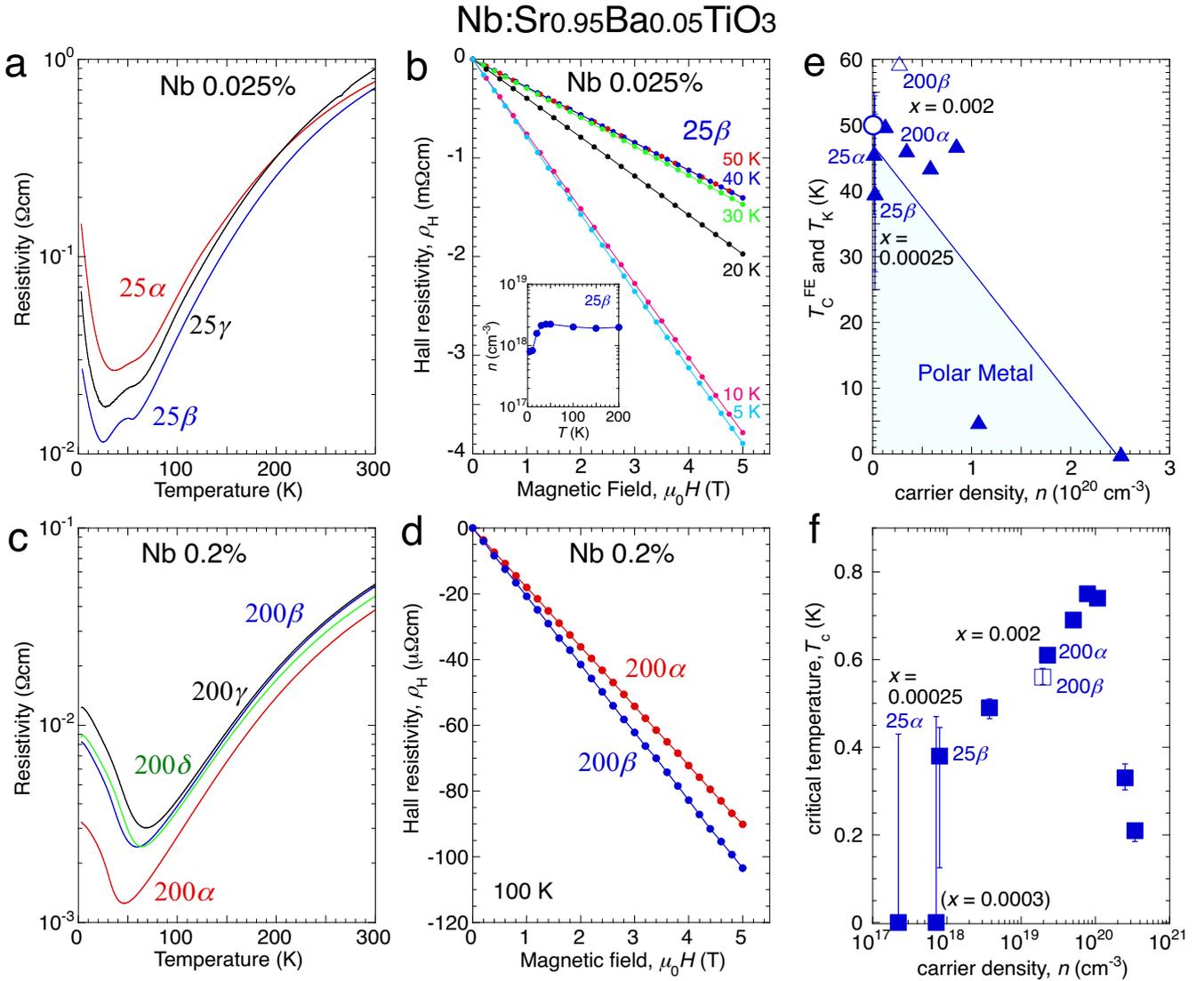

**Supplementary Figure 7 | Variation of the carrier density *n*, resistance anomaly temperature $T_K$, and the superconducting transition temperature $T_c$.** **a**, Resistivity vs temperature for three different single crystals (denoted by 25α, 25β, and 25γ) of Nb:$Sr_{0.95}Ba_{0.05}TiO_3$ with Nb contents of 0.025%. The itinerant electrons per unit formula at 100 K deduced from the Hall effect are 0.00008 for 25α and 0.00009 for 25β. We have not measured the Hall effect for 25γ. The samples 25β and 25γ show two anomaly temperatures $T_K$'s, whereas for the sample 25α, the two anomalies seem to be merged. **b**, Hall resistivity vs magnetic field for the 25β sample. The carrier density decreases below around 20 K, as shown in the inset. **c**, Resistivity vs temperature for four different single crystals (denoted by 200α, 200β, 200γ, and 200δ) of Nb:$Sr_{0.95}Ba_{0.05}TiO_3$ with Nb contents of 0.2%. The number of itinerant electrons per unit formula at room temperature is 0.002 for 200α and 0.0018 for 200β. All the samples show a single anomaly, which deviates from 40 K to 70 K. **d**, Hall resistivity vs magnetic field for the 200α and 200β samples at 100 K. **e**, Resistance anomaly temperature $T_K$ vs carrier density *n* at $T_K$ or 5 K. **f**, Superconducting transition temperature $T_c$ vs carrier density *n* at 5 K.